\documentclass[fleqn]{aa}
\pdfoutput=1

\usepackage[varg]{txfonts}
\makeatletter
\renewcommand*\aa@pageof{, page \thepage{} of \pageref*{LastPage}}
\makeatother
\usepackage[utf8]{inputenc}
\usepackage[english]{babel}

\usepackage{natbib,twoopt}
\usepackage[pdfborderstyle={/S/U/W 0}, linktoc=page, bookmarks=true, breaklinks=true, colorlinks=true, linkcolor=blue, urlcolor=blue, citecolor=blue]{hyperref} 
\bibpunct{(}{)}{;}{a}{}{,} 
\makeatletter
\newcommandtwoopt{\citeads}[3][][]{\href{http://adsabs.harvard.edu/abs/#3}%
{\def\hyper@linkstart##1##2{}%
\let\hyper@linkend\@empty\citealp[#1][#2]{#3}}}
\newcommandtwoopt{\citepads}[3][][]{\href{http://adsabs.harvard.edu/abs/#3}%
{\def\hyper@linkstart##1##2{}%
\let\hyper@linkend\@empty\citep[#1][#2]{#3}}}
\newcommandtwoopt{\citetads}[3][][]{\href{http://adsabs.harvard.edu/abs/#3}%
{\def\hyper@linkstart##1##2{}%
\let\hyper@linkend\@empty\citet[#1][#2]{#3}}}
\newcommandtwoopt{\citeyearads}[3][][]%
{\href{http://adsabs.harvard.edu/abs/#3}
{\def\hyper@linkstart##1##2{}%
\let\hyper@linkend\@empty\citeyear[#1][#2]{#3}}}
\makeatother

\usepackage{textgreek}
\usepackage{graphicx}	
\usepackage{tablefootnote}
\usepackage{amsmath, bm}	
\usepackage{amssymb}	
\usepackage{url}
\usepackage[separate-uncertainty]{siunitx}
\usepackage{multicol,multirow, makecell}
\usepackage{enumitem}
\setlist{  
  listparindent=\parindent,
  parsep=0pt,
}

\defcitealias{Asgari21}{A21}

\usepackage{xcolor}
\usepackage[normalem]{ulem}

\begin{document}

\title{KiDS-1000: cosmic shear with enhanced redshift calibration}

\author{
  J.~L.~van den Busch\inst{\ref{inst:RUB},\ref{inst:AIfA}}
  \and
  A.~H.~Wright\inst{\ref{inst:RUB}}
  \and
  H.~Hildebrandt\inst{\ref{inst:RUB}}
  \and
  M.~Bilicki\inst{\ref{inst:CTP}}
  \and
  M.~Asgari\inst{\ref{inst:ROE},\ref{inst:UHull}}
  \and
  S.~Joudaki\inst{\ref{inst:waterloo1},\ref{inst:waterloo2}}
  \and
  C.~Blake\inst{\ref{inst:swin}}
  \and
  C.~Heymans\inst{\ref{inst:ROE},\ref{inst:RUB}}
  \and
  A.~Kannawadi\inst{\ref{inst:APrince}}
  \and
  H.~Y.~Shan\inst{\ref{inst:SHAO},\ref{inst:UCAS}}
  \and
  T.~Tröster\inst{\ref{inst:ROE}}
}

\institute{
  Ruhr-University Bochum, Astronomical Institute, German Centre for Cosmological Lensing, Universitätsstr. 150, 44801, Bochum, Germany, \email{jlvdb@astro.rub.de}\label{inst:RUB}
  \and
  Argelander-Institut für Astronomie, Universität Bonn, Auf dem Hügel 71, 53121 Bonn, Germany\label{inst:AIfA}
  \and
  Center for Theoretical Physics, Polish Academy of Sciences, al. Lotników 32/46, 02-668, Warsaw, Poland\label{inst:CTP}
  \and
  Institute for Astronomy, University of Edinburgh, Royal Observatory, Blackford Hill, Edinburgh, EH9 3HJ, U.K.\label{inst:ROE}
  \and
  E. A. Milne Centre, University of Hull, Cottingham Road, Hull, HU6 7RX, UK\label{inst:UHull}
  \and
  Waterloo Centre for Astrophysics, University of Waterloo, 200 University Ave W, Waterloo, ON N2L 3G1, Canada\label{inst:waterloo1}
  \and
  Department of Physics and Astronomy, University of Waterloo, 200 University Ave W, Waterloo, ON N2L 3G1, Canada\label{inst:waterloo2}
  \and
  Centre for Astrophysics \& Supercomputing, Swinburne University of Technology, P.O. Box 218, Hawthorn, VIC 3122, Australia\label{inst:swin}
  \and
  Department of Astrophysical Sciences, Princeton University, 4 Ivy Lane, Princeton, NJ 08544, USA\label{inst:APrince}
  \and
  Shanghai Astronomical Observatory (SHAO), Nandan Road 80, Shanghai 200030, China\label{inst:SHAO}
  \and
  University of Chinese Academy of Sciences, Beijing 100049, China\label{inst:UCAS}
}

\date{Received 25 August 2021 / Accepted 4 April 2022}

\abstract{
  We present a cosmic shear analysis with an improved redshift calibration for the fourth data release of the Kilo-Degree Survey (KiDS-1000) using self-organising maps (SOMs). Compared to the previous analysis of the KiDS-1000 data, we expand the redshift calibration sample to more than twice its size, now consisting of data of 17 spectroscopic redshift campaigns, and significantly extending the fraction of KiDS galaxies we are able to calibrate with our SOM redshift methodology. We then enhance the calibration sample with precision photometric redshifts from COSMOS2015 and the Physics of the Accelerated Universe Survey (PAUS), allowing us to fill gaps in the spectroscopic coverage of the KiDS data. Finally we perform a Complete Orthogonal Sets of E/B-Integrals (COSEBIs) cosmic shear analysis of the newly calibrated KiDS sample. We find $S_8 = 0.748_{-0.025}^{+0.021}$, which is in good agreement with previous KiDS studies and increases the tension with measurements of the cosmic microwave background to 3.4σ. We repeat the redshift calibration with different subsets of the full calibration sample and obtain, in all cases, agreement within at most 0.5σ in $S_8$ compared to our fiducial analysis. Including additional photometric redshifts allows us to calibrate an additional 6 \% of the source galaxy sample. Even though further systematic testing with simulated data is necessary to quantify the impact of redshift outliers, precision photometric redshifts can be beneficial at high redshifts and to mitigate selection effects commonly found in spectroscopically selected calibration samples.
}


\maketitle



\section{Introduction}

Over the past decade, gravitational lensing \citep{Bartelmann01} has emerged as one of the most powerful tools to study gravity and the dark sectors of the Universe, dark matter and dark energy, through the impact of these components on the density fluctuations of matter and their evolution with cosmic time \citep{Peacock06}. In the limit of weak lensing, massive structures along the line of sight imprint a subtle shearing on the shapes of distant galaxies. This signal can be extracted by statistically analysing the ellipticity of galaxy images in large surveys \citep{Refregier03}. These cosmic shear surveys \citep{Kilbinger15} face the challenge that they must accurately reconstruct the galaxy redshift distribution in order to interpret the cosmological signal correctly. Even small biases in the first moment of the redshift distribution may introduce significant biases in the recovered cosmological parameters \citep[e.g.][]{Huterer06,Ma06}. There exists tension between constraints from cosmic shear and the cosmic microwave background, first seen between the Canada-France-Hawaii Telescope Lensing Survey \citep[CFHTLenS,][]{Heymans13,MacCrann15,Joudaki17} and Planck \citep{Planck14}, but for example also between the Kilo-Degree Survey \citep[KiDS,][]{Kuijken15} and Planck legacy \citep{Planck20}, recently for KiDS-1000 \citep{Asgari21}. The most recent cosmic shear results from the Dark Energy Survey \citep[DES,][]{Flaugher15} are very similar to KiDS-1000 \citep{Amon21,Secco21} albeit at lower statistical tension with Planck\footnote{Conclusions over the degree of tension differ primarily owing to the way tension is quantified and different prior choices for the neutrino mass in the Planck re-analysis.}. In the light of these repeatedly reported tensions, redshift calibration has come under scrutiny as one of the systematics for cosmic shear experiments \citep[e.g.][]{Joudaki20}.

Due to the statistical nature of the shear measurements, current generation (stage-III) cosmic shear surveys, such as KiDS, DES, and the Hyper Suprime-Cam Subaru Strategic Program \citep[HSC,][]{Aihara18}, rely on imaging of tens of millions of galaxies for which spectroscopic redshifts cannot be measured directly. Instead, galaxy redshifts are determined with secondary redshift estimates, the most notable ones are direct calibration with spectroscopic training samples \citep[e.g.][]{Lima08,Hildebrandt17,Hildebrandt20,Buchs19,Wright20a}, clustering redshifts (which infer redshift distributions by exploiting the gravitational clustering of galaxies at similar redshifts, e.g. \citealt{Newman08,Matthews10,Schmidt13,Menard13,vandenBusch20,Hildebrandt21,Gatti22}) and methods that make use of a combination of both these approaches \citep{Sanchez18,Alarcon20a,Myles21}.

The redshift calibration of the fourth data-release of KiDS \citep{Kuijken19,Hildebrandt21} relies on an implementation of the direct calibration that utilises a self-organising map \citep[SOM,][]{Kohonen82,Wright20a} based on work by \citet{Masters16}. The fundamental principle of this method is to re-weight a spectroscopic reference sample such that it is representative of a photometric dataset with an unknown redshift distribution. The weighted redshift distribution of the reference sample is then a direct estimate of the unknown distribution. Additionally, the SOM method allows for the removal of galaxies from the photometric dataset for which no similar galaxies exist in the reference sample. Their inclusion would otherwise bias the estimated redshift distribution. We call the subset of the remaining, well represented galaxies the `gold sample'. In this work we explore the redshift calibration of the KiDS data with a significantly enhanced reference sample that is composed of a variety of spectroscopic redshift campaigns and precision photometric redshifts. This allows us to expand the KiDS gold sample and calibrate redshifts in regions of the colour space that are difficult to access by direct spectroscopy. We then study how these additional calibration data influence our ability to calibrate the redshifts of the source sample and how selection effects and changes in the calibration propagate to cosmological constraints. We compare our results to the original KiDS-1000 cosmic shear analysis by \citet{Asgari21}.

This paper is structured as follows: in Sect.~\ref{sec:data} we describe the KiDS data and the redshift calibration sample (further details on this compilation in App.~\ref{app:compilation}), and in Sect.~\ref{sec:redshifts} and~\ref{sec:cosmo} we present the SOM redshift calibration and our cosmic shear analysis methods. We present and discuss the newly calibrated gold samples and cosmological constraints in Sect.~\ref{sec:results} and~Sect.~\ref{sec:discussion}. Finally we conclude and summarise in Sect.~\ref{sec:conclusion}.


\section{Data}
\label{sec:data}

This paper explores redshift calibration of the KiDS cosmic shear weak lensing sample (Sect.~\ref{sec:data:k1000}) with increasingly deep redshift calibration catalogues. Our fiducial analysis relies exclusively on spectroscopic data which is compiled from a variety of different spectroscopic surveys (Sect.~\ref{sec:data:spec} and App.~\ref{app:compilation}). We then add less accurate redshift estimates derived from narrowband photometry from the Physics of the Accelerated Universe Survey \citep[PAUS,][]{Padilla19} and finally from medium- and broadband photometry from COSMOS2015 (Sect.~\ref{sec:data:phot}). This approach is similar to the construction of the calibration sample for the DES Y3 redshift calibration \citep{Myles21}, however the KiDS lensing and redshift calibration samples are both covered by the same nine bands which avoids mapping different photometries via a transfer function.

\subsection{KiDS-1000 photometric data}
\label{sec:data:k1000}

The Kilo-Degree Survey \citep[KiDS,][]{Kuijken15,deJong15,deJong17,Kuijken19} is a public European Southern Observatory (ESO) survey that has been designed particularly with weak gravitational lensing applications in mind. The complete survey will deliver about \SI{1350}{deg^2} of $ugri$ imaging split into an equatorial and a southern field. Combined with $ZYJHK_{\rm s}$ imaging from its companion infra-red survey, the VISTA Kilo-Degree Infrared Galaxy Survey \citep[VIKING][]{Edge13,Venemans15}, this constitutes a nine-band, matched-depth data-set with primary imaging in the $r$-band, observed at a mean seeing of \SI{0.7}{\arcsec}. This work is based on the fourth data release of KiDS which covers \SI{1006}{deg^2}. The weak lensing source catalogue \citep[KiDS-1000,][]{Giblin21} is divided into five tomographic redshift bins, based on nine-band photometric redshifts (four bins with $\Delta Z_{\rm B} = 0.2$, starting from $Z_{\rm B} = 0.1$ and a fifth bin at $0.9 < Z_{\rm B} \leq 1.2$) computed with \texttt{BPZ} \citep[Bayesian Photometric Redshift,][]{Benitez00}. It contains all objects with non-zero shear measurement weights obtained from {\it lens}fit \citep{Miller07,FenechConti17}, which effectively selects objects with $r$-band magnitudes between $20 \leq r \leq 25$.

In addition to the main survey imaging there are observations of six fields dedicated primarily to redshift calibration. These `KiDZ' fields cover approximately \SI{1}{deg^2} each and target areas of the sky also observed by different spectroscopic campaigns, which are summarised in Sect.~\ref{sec:data:spec}. Just like the main survey, KiDZ provides KiDS+VIKING 9-band imaging\footnote{In case of the COSMOS field we instead use existing data in the CFHT (Canada France Hawaii Telescope) $z$-band which is sufficiently similar to the VISTA InfraRed CAMera (VIRCAM) $Z$-band at the Visible and Infrared Survey Telescope for Astronomy (VISTA) given the photometric and redshift calibration uncertainties.} which reaches or exceeds the depth of the main survey. In the latter case we homogenise the data depth by applying Gaussian noise to obtain matched photometry.

\subsection{Spectroscopic data for calibration}
\label{sec:data:spec}

The most important spectroscopic campaigns that overlap with the six KiDZ fields are zCOSMOS \citep{Lilly09}, VVDS \citep[VIMOS VLT Deep Survey,][]{LeFevre05,LeFevre13}, DEEP2 \citep{Newman13}, the GAMA \citep[Galaxy And Mass Assembly,][]{Driver11} deep field G15deep \citep{Kafle18,Driver22}, and a compilation of spectra covering the Chandra Deep Field South (CDF-S, App.~\ref{app:compilation}). This data has been used in previous KiDS redshift calibration works \citep{Wright20a,Hildebrandt21}.

\begin{figure}
    \centering
    \includegraphics[width=0.87\columnwidth]{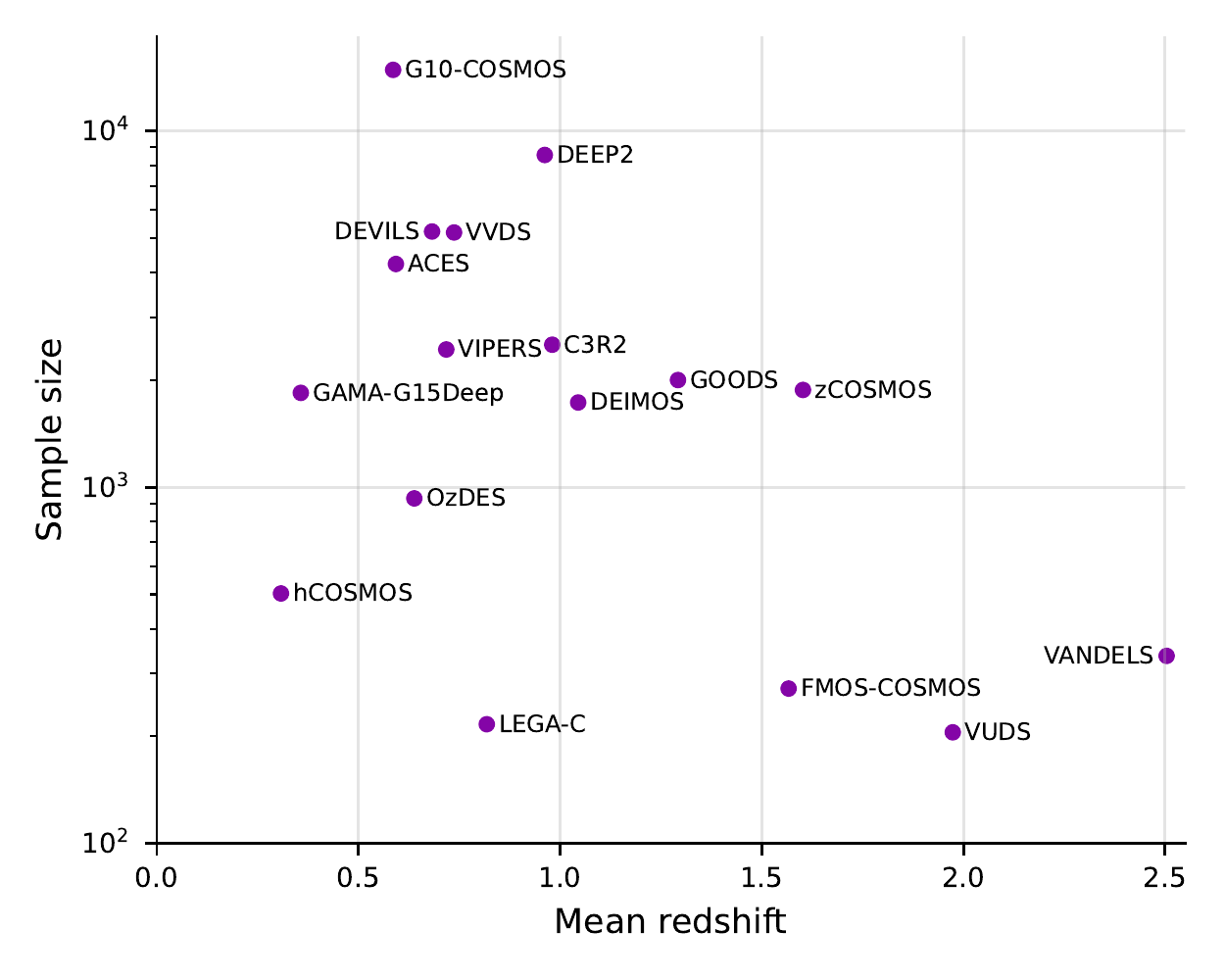}
    \caption{
        Sample size and mean redshift of the different surveys that are part of our spectroscopic compilation after matching to their counterparts in KiDS imaging. Objects with redshifts from multiple sources are assigned to the survey with the most reliable redshift estimate.}
    \label{fig:samples}
\end{figure}

We extend this compilation by adding data from C3R2 \citep[Complete Calibration of the Colour-Redshift Relation,][]{Masters17,Masters19,Euclid20,Stanford21}, DEVILS \citep[Deep Extragalactic Visible Legacy Survey,][]{Davies18}, VIPERS \citep[VIMOS\footnote{VIsible Multi-Object Spectrograph} Public Extragalactic Redshift Survey,][]{Scodeggio18} and a variety of spectroscopic campaigns that target the CDF-S and COSMOS fields which are detailed in App.~\ref{app:compilation}. We also revise the selection of sources included for calibration by removing duplicates, both from spatial overlap as well as within the datasets, and by homogenising redshift quality flags based on the original information in the input samples. If, for a given source, there are redshifts from different surveys available, we assign the most reliable measurement based on a specific `hierarchy' of surveys (see App.~\ref{app:compilation} for details). For objects with multiple spectroscopic measurements within a particular survey, we either take the redshift with the highest quality flag or, if various entries for the same source have the same quality flag and the reported redshifts differ by no more than 0.005, we take the average. However, if the reported redshift differences exceed this threshold, we exclude such a source from the compilation. We restrict the selection to objects with high quality spectroscopic redshifts (approximately corresponding to $\geq \SI{95}{\percent}$ confidence or redshift quality code ${\rm nQ} \geq 3$). Fig.~\ref{fig:samples} compares the number of galaxies and their mean redshift for all samples that enter the spectroscopic compilation. These values apply after removing duplicates between overlapping catalogues and only for those objects with photometric coverage in KiDZ (Sect.~\ref{sec:data:k1000}).

\subsection{Photometric data for calibration}
\label{sec:data:phot}

The success rate of spectroscopically determined redshifts is very different from (typically flux-limited) imaging data. Therefore, it is very difficult to obtain a spectroscopic calibration sample that is representative of photometric data in magnitude-\slash colour-space, especially at faint magnitudes. Instead, we additionally include galaxy samples with high quality photometric redshifts to achieve a greater overall coverage of the KiDS data by the calibration sample which is beneficial for our redshift calibration technique of choice (Sect.~\ref{sec:SOM}).

\subsubsection{COSMOS2015}

The COSMOS2015 catalogue \citep{Laigle16} constitutes a sample of about half a million galaxies in the COSMOS field with precision photometric redshifts\footnote{We use the median of the photo-$z$ likelihood distribution (\texttt{PHOTOZ} column in the catalogue).} derived from up to 30 photometric bands, ranging from near ultra-violet to mid infrared, including 14 medium and narrow band filters. This sample extends to higher redshifts ($z_{\rm max} \approx 6$) and fainter magnitudes than our spectroscopic compilation, but at the cost of less secure redshift estimates with an outlier fraction ranging from \SI{0.5}{\percent} at low redshifts to \SI{13.2}{\percent} for $3 < z < 6$ \citep{Laigle16}.

\subsubsection{PAUS}

The PAUS photometric redshift sample\footnote{available at \url{cosmohub.pic.es}} \citep{Alarcon20b} is a combination of 26 optical and near-infrared bands from COSMOS2015 that are matched against observations of the COSMOS field in 40 narrow band filters by the PAU survey. These PAU filters sample the optical regime between \SIrange{450}{850}{nm} at a bandwidth of $\Delta \lambda = \SI{12.5}{nm}$ \citep{Padilla19} and the combined photometric catalogue is limited to $i_{\rm AB} < 23$. Due to the relatively high spectral resolution of the dataset a new Bayesian spectral energy distribution (SED) fitting technique is required that accounts for individual emission lines \citep{Alarcon20b}. This allows the PAUS photo-$z$ to achieve a $3\times$ ($1.7\times$) lower photo-$z$ scatter at the bright (faint) end of the magnitude distribution and marginally smaller outlier fractions compared to the original COSMOS2015 photo-$z$ at $i_{\rm AB} < 23$. The scaled photo-$z$ bias is very low and has a $|\text{median}(\Delta_z)| < 0.001$ over the whole redshift range of the PAUS sample. Therefore this sample positions itself right between the spectroscopic data and COSMOS2015 in terms of completeness and redshift precision.

\subsection{Combined calibration sample}
\label{sec:compilation}

In this work we select data from the full COSMOS2015 photometric catalogue and combine this data hierarchically with PAUS and the spectroscopic data. Finally we match this unified catalogue to the KiDZ imaging to form our redshift calibration sample.

We prepare the full COSMOS2015 photometric catalogue similar to \citet{Laigle16}, that is, we select only those sources which fall into the intersection of the footprint of the COSMOS field (${\tt flag\_cosmos} = 1$) and the UltraVISTA observations (${\tt flag\_hjmcc} = 0$), which provides essential IR spectral coverage. We exclude data from saturated areas ($\mathtt{flag\_peter} = 0$) and additionally remove objects that are classified to be most likely stars (${\tt type} \neq 1$) or have no photo-$z$ estimate. This selection yields about half a million objects.

About 40,000 sources of the PAUS sample are, by design, matched against COSMOS2015 and therefore require no further preparation. Therefore we can directly combine the spectroscopic compilation, PAUS, and the subset of COSMOS2015 by matching objects within \SI{1}{\arcsec}. We maintain a hierarchy to ensure that we always choose the most reliable redshift estimate available: spec-$z$ supersede PAUS photo-$z$ which supersede the COSMOS2015 photo-$z$. Finally, we assign 9-band KiDS magnitudes to this compilation by matching against the KiDZ data, again within \SI{1}{\arcsec}. This combination of spec-$z$ (in all KiDZ fields), photo-$z$ (only in COSMOS) and KiDS imaging represents our full redshift calibration sample.

The method to combine the two photometric redshift samples with our spectroscopic compilation in an hierarchical manner is very similar to the approach taken for the redshift calibration of the DES Y3 data \citep{Myles21}. There are, however, two key differences to their approach. First, our compilation of spectroscopic redshifts covers a much wider range of the colour-redshift space than their selection of spectra. This allows us to construct more representative calibration samples that consist purely of spectroscopic redshifts, photometric redshifts, or a combination thereof. Second, the primary KiDS imaging data and the calibration data are observed in all nine photometric bands at a comparable depth, which simplifies the mapping from galaxy colour to redshift significantly (see Sect.~\ref{sec:redshifts}).

\subsubsection{Primary compilations}
\label{sec:comp_primary}

\begin{figure}
    \centering
    \includegraphics[width=\columnwidth]{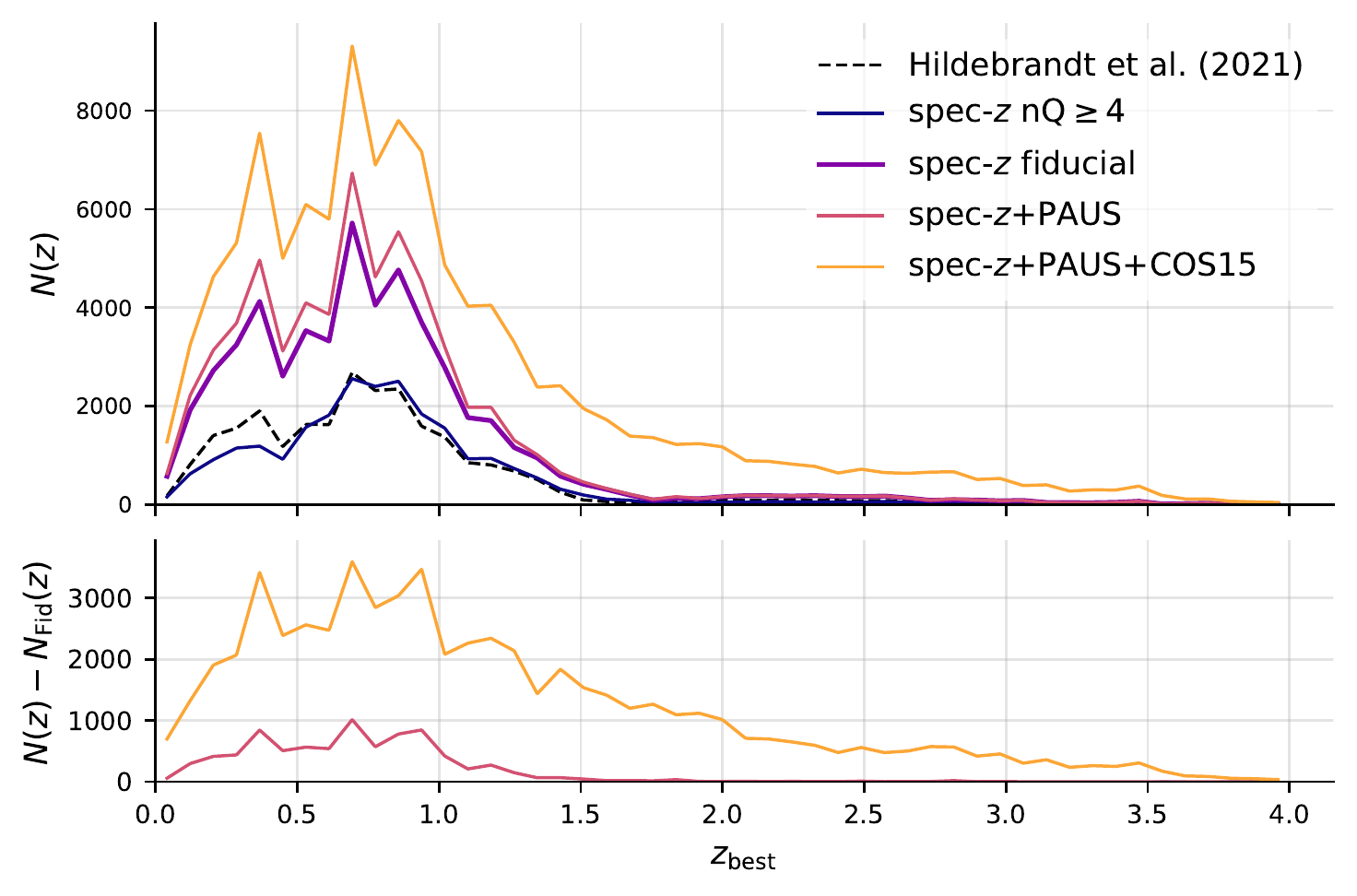}
    \caption{
        Redshift distribution of different calibration sample subsets (coloured, solid lines) and the one used by \citet{Hildebrandt21} for KiDS-1000 (dashed black line). The bottom panel shows the excess of sources with photometric redshifts contributed by PAUS and PAUS + COSMOS2015 compared to the fiducial spectroscopic sample.}
    \label{fig:nz_calib}
\end{figure}

From this heterogeneous sample with redshift estimates from very different sources we select three subsets, each with a higher redshift precision but lower completeness:
\begin{itemize}
    \item the full compilation (to which we refer as {\sl \mbox{spec-$z$}+\allowbreak PAUS+\allowbreak COS15}),
    \item objects with either spectroscopic redshifts and/or PAUS photo-$z$ ({\sl \mbox{spec-$z$}+\allowbreak PAUS}), and
    \item our fiducial sample containing only those objects that have spectroscopic redshifts ({\sl \mbox{spec-$z$} fiducial}).
\end{itemize}
The main properties and redshift distributions of these three primary compilations are summarised in Table~\ref{tab:samples} and Fig.~\ref{fig:nz_calib}.

The fiducial sample is already about twice as large as the calibration sample used previously by \citet{Hildebrandt21} to calibrate the KiDS-1000 redshifts. Of the additional spectra we consider DEVILS and C3R2, the latter of which is designed to target regions of the galaxy colour-space with currently little spectroscopic coverage, to be the most important contributions. Similarly {\sl \mbox{spec-$z$} fiducial} already contains about \SI{66}{\percent} of the matched PAUS sources, of which the majority has redshift $z < 1$. Due to its limited depth, the {\sl \mbox{spec-$z$}+\allowbreak PAUS} sample presents only a small improvement over the fiducial case. The COSMOS2015 data, on the contrary, nearly doubles the compilation to its final size of about $112,000$ objects. Due to the significantly higher depth of the COSMOS2015 photo-$z$, the fraction of sources with $z > 1$ nearly triples, pushing the mean redshift to $\langle z \rangle \approx 1.0$. Nevertheless this comes at the cost of a lower redshift accuracy compared to the rest of the sample.

\begin{table}
    \centering
    \caption{
        Number counts and mean redshifts of the original KiDS-1000 redshift calibration sample and different subsets of the new redshift compilation.}
    \label{tab:samples}
    \begin{tabular}{lcc}
\hline\hline
Compilation &            Count & $\langle z \rangle$ \\
\hline
\citet{Hildebrandt21}         &  $\hphantom{0}25\,373$ &             $0.796$ \\
spec-$z$ fiducial             &  $\hphantom{0}52\,911$ &             $0.788$ \\
spec-$z$+PAUS                 &  $\hphantom{0}61\,163$ &             $0.776$ \\
spec-$z$+PAUS+COS15           &  $\hphantom{}112\,400$ &             $1.002$ \\
spec-$z$ $\mathrm{nQ} \geq 4$ &  $\hphantom{0}24\,117$ &             $0.832$ \\
only PAUS                     &  $\hphantom{0}24\,229$ &             $0.640$ \\
only COS15                    &  $\hphantom{0}80\,632$ &             $1.081$ \\
only PAUS+COS15               &  $\hphantom{0}80\,635$ &             $1.084$ \\
\hline
\end{tabular}

    \tablefoot{
        The full compilation is represented as {\sl \mbox{spec-$z$}+\allowbreak PAUS+\allowbreak COS15}. Values apply after matching with the KiDZ imaging and removing duplicates in the spectroscopic data.}
\end{table}

\subsubsection{Secondary compilations}
\label{sec:comp_secondary}

In addition to the three primary compilations we also consider a subset that is restricted to only the most secure spectroscopic redshifts. This {\sl \mbox{spec-$z$} \mbox{nQ $\geq 4$}} sample is, due to the large fraction of shared spectra, closest to the one of \citet{Hildebrandt21} except that it lacks some low redshift sources (Fig.~\ref{fig:nz_calib})

Finally, we create three subsets of the full redshift compilation that rely purely on photometric redshift estimates. We achieve this by recompiling the redshift compilation according to Sect.~\ref{sec:compilation} but omitting all spectroscopic redshifts, therefore maintaining the usual hierarchy of PAUS and COSMOS2015 photo-$z$. These are
\begin{itemize}
    \item only objects from the PAU survey ({\sl only-PAUS}),
    \item all objects with photo-$z$ ({\sl only-PAUS+\allowbreak COS15}), and
    \item the pure COSMOS2015 subset ({\sl only-COS15}, also discarding PAUS photo-$z$ from the stack).
\end{itemize}
Since the PAUS sample is essentially a subset of the COSMOS2015 catalogue, the latter two samples are almost identical except that for \SI{30}{\percent} of the sources the photo-$z$ are augmented by the PAUS data (see Table~\ref{tab:samples}). The PAUS sample is about half the size of the fiducial spectroscopic compilation and, while achieving a higher completeness at $i_{\rm AB} < 23$, lacks many important faint, high redshift objects.


\section{Redshift calibration with self-organising maps}
\label{sec:redshifts}

A self-organising map \citep[SOM,][]{Kohonen82} is a very powerful tool that allows us to calibrate the redshift distribution of the KiDS-1000 lensing sample using the redshift compilations defined in the previous section. We adopt the SOM methodology of \cite{Wright20a} which additionally provides a metric to select only those parts of the KiDS colour space in which we can reliably map out the colour-redshift relation.

\subsection{SOM redshift calibration methodology}
\label{sec:SOM}

The basic idea of the SOM methodology dates back to \citet{Lima08} who introduced a redshift calibration strategy built on the assumption that two galaxy samples with the same colour-space distribution follow the same redshift distribution. Therefore they suggest to derive the unknown redshift distribution $N(z)$ of a photometric galaxy sample from a calibration sample with accurate, preferentially spectroscopic redshifts $N^{\rm cal}(z)$ that is constructed such that it is representative of the photometric sample. This method is called `direct calibration' (DIR). In practice, however, such a calibration sample has typically a substantially different selection function. Therefore \citet{Lima08} propose a re-weighting scheme to match the calibration to the photometric sample by computing the ratio of the local galaxy density of both samples in the high-dimensional colour-space spanned by the photometric observations. This can be achieved for example by counting neighbours in a fixed volume around a point in the colour-space or by computing the volume occupied by a fixed number of nearest neighbours. Provided that both samples initially cover the same volume of the colour-space this method should recover the true redshift distribution, even in the presence of colour-redshift degeneracies.

This method is still susceptible in particular to selection biases and incompleteness introduced by spectroscopic targeting strategies and success rates \citep[e.g.][]{Gruen17,Hartley20}. Recent work by \citet{Wright20a} has shown that this can be alleviated by performing additional cleaning and selections (quality control, see Sect.~\ref{sec:SOMkids}) on the unknown sample, creating a `gold sample' containing only galaxies of the photometric sample that are sufficiently represented by the calibration sample. They implement this by training a SOM on the colour-space of the calibration sample and then parse the photometric sample into the same cells. Cells that are not occupied by objects from both samples are rejected, effectively removing those critical parts of the colour-space. They improve the cleaning procedure by applying hierarchical clustering on the SOM to find groups of cells with similar photometric properties instead of filtering individual cells. This allows a more fine-grained trade-off between the number of photometric sources rejected due to partitioning of the high-dimensional colour-space and the bias introduced by misrepresentation of the gold sample.

Finally, they compute the DIR weight for each of the $n$ SOM groupings $\mathcal{G} = \{g_1, \dots, g_n\}$ which is the ratio of calibration-to-gold sample objects. They obtain the redshift distribution of the gold sample
\begin{equation}
    N(z) = \sum_{g \,\in\, \mathcal{G}} N_g^{\rm cal}(z) \, \frac{N_{g,{\rm tot}}^{\rm gold}}{N_{g,{\rm tot}}^{\rm cal}}
    \label{eq:som_weight}
\end{equation}
by calculating the DIR-weighted sum of the redshift distributions $N_g^{\rm cal}(z)$ of the calibration sample in each SOM grouping. $N_{g,{\rm tot}}^{\rm cal}$ and $N_{g,{\rm tot}}^{\rm gold}$ are the total number of calibration sample and gold sample objects of group $g$, respectively.

\subsection{Application to KiDS-1000}
\label{sec:SOMkids}

For our analysis we largely follow \citet{Wright20a} and train a SOM with $101\times101$ hexagonal cells and periodic boundaries on the full calibration sample ({\sl \mbox{spec-$z$}+\allowbreak PAUS+\allowbreak COS15}, see Sect.~\ref{sec:comp_primary}). The input features are the matched KiDS $r$-band magnitudes and all 36 possible KiDS-colours that can be formed from the $ugriZYJHK_{\rm s}$ imaging. Next, we divide the calibration and the KiDS-1000 source sample into the five tomographic bins and parse both samples into the SOM cells. We then run the hierarchical clustering for which we use the same number of clusters per bin (4000, 2200, 2800, 4200, and 2000) as \citet{Wright20a} since these numbers were calibrated using simulations\footnote{These simulations are tailored to fit the KiDS imaging data and are based on the MICE2 simulation \citep{Fosalba15a,Fosalba15b,Crocce15,Carretero15,Hoffmann15}.} \citep{vandenBusch20}. Even though each gold sample has a different optimal number of clusters, simulating the new redshift compilation and including realistic photo-$z$ errors is beyond the scope of this work.

We are using the same SOM for the remaining calibration samples defined in Sect.~\ref{sec:compilation} and simply parse the corresponding subset of the full calibration sample back into the SOM before running the hierarchical clustering. For each of these calibration samples we apply a final cleaning step to the SOM groupings by defining a quality cut
\begin{equation}
    | \langle z_{\rm cal} \rangle - \langle z_{\rm B} \rangle | > 5 \sigma_{\rm mad} \, ,
    \label{eq:qc}
\end{equation}
where $\sigma_{\rm mad} = \mathrm{nMAD}\left( \langle z_{\rm cal} \rangle - \langle z_{\rm B} \rangle \right)$ is the normalised median absolute deviation from the median, where the normalisation ensures that the nMAD reproduces the traditional standard deviation in the limit of Gaussian noise. This selection rejects clusters of SOM cells in which the mean calibration sample redshift $\langle z_{\rm cal} \rangle$ and the mean KiDS photometric redshifts $\langle z_{\rm B} \rangle$ catastrophically disagree. \citet{Wright20a} found that this additional cleaning significantly reduces the SOM redshift bias while the impact on the number density is small and does not exceed a few percent. The rejection threshold of $\sigma_{\rm mad} \approx 0.12$ is calculated for the {\sl \mbox{spec-$z$} fiducial} case and is applied to all other samples. This choice is motivated by the fact that this value is very close to the one calibrated with mock data for KiDS-1000 by \citep{Hildebrandt21}, whereas $\sigma_{\rm mad}$ would be twice as large if we were calculating this threshold based on {\sl \mbox{spec-$z$}+\allowbreak PAUS+\allowbreak COS15}. One reason for this difference in $\sigma_{\rm mad}$ is that the COSMOS2015 data allow the inclusion of additional populations of faint galaxies for which the calibration sample reference redshifts and the KiDS photo-$z$ are more likely discrepant, increasing the spread of the distribution of $\langle z_{\rm cal} \rangle - \langle z_{\rm B} \rangle$. We discuss this effect further in Sect.~\ref{sec:discuss_gold} and App.~\ref{app:SOMqc}.

This final selection step defines our gold sample for which we compute the redshift distributions according to Eq.~(\ref{eq:som_weight}). Since we require weighted redshift distributions for our cosmological analysis we substitute $N_{g,{\rm tot}}^{\rm gold}$ by $W_{g,{\rm tot}}^{\rm gold} = \sum_{i \in g} w_i$ which is the sum over the individual galaxy weights $w_i$ from shape measurements in the SOM group $g$.

\subsection{Clustering redshifts}

There is one key difference to the calibration methodology of \citet{Hildebrandt21} which is that we choose to omit the clustering redshift analysis in this work. While this choice limits our ability to validate the redshift distributions of our new gold samples, the SOM method is our fiducial calibration method and is therefore the focus of this work. In addition to that, the newly included calibration data does not increase the spatial overlap with KiDS significantly and is, due to its inhomogeneity, difficult to administer in a cross-correlation analysis. We leave this validation and joint analysis with the clustering redshifts to future work.


\section{Cosmological analysis}
\label{sec:cosmo}

In this section we summarise our cosmic shear analysis pipeline which we adopt from \citet[\citetalias{Asgari21} hereafter]{Asgari21}.

\subsection{Cosmic shear}

The primary observable of cosmic shear are the shear two-point correlation functions \citep[2PCFs,][]{Kaiser92}
\begin{equation}
    \xi_\pm(\theta) = \langle \gamma_{\rm t} \gamma_{\rm t} \rangle(\theta) \pm \langle \gamma_\times \gamma_\times \rangle(\theta) \, ,
\end{equation}
where $\gamma_{\rm t}$ and $\gamma_\times$ are the tangential and the cross component of the shear, defined with respect to the line connecting a pair of galaxies \citep[see e.g.][]{Bartelmann01}. We use a weighted estimator for the shear correlations $\xi_\pm$ as a function of the separation angle $\theta$ between two tomographic redshift bins $i$ and $j$:
\begin{equation}
    \hat\xi_\pm^{(ij)}(\bar \theta) = \frac{\sum_{ab} w_a w_b \left[ \epsilon_{{\rm t},a}^{\rm obs} \epsilon_{{\rm t},b}^{\rm obs} \pm \epsilon_{\times,a}^{\rm obs} \epsilon_{\times,b}^{\rm obs} \right] \Delta_{ab}^{(ij)}(\bar \theta)}{\sum_{ab} w_a w_b (1 + \bar m_a) (1 + \bar m_b) \, \Delta_{ab}^{(ij)}(\bar \theta)}
    \label{eq:shear_estimator}
\end{equation}
Here, $\Delta_{ab}^{(ij)}(\bar \theta)$ is a function that expresses whether a pair of galaxies, $a$ and $b$, falls into an angular bin labelled by $\bar\theta$. Each galaxy has a weight $w$ and measured ellipticities $\epsilon_{\rm t}^{\rm obs}$ and $\epsilon_\times^{\rm obs}$. The denominator applies the multiplicative shear bias $m$, which corrects the measured shear to match the true galaxy shear.\footnote{In practice we do not apply the $m$-bias per galaxy in Eq.~(\ref{eq:shear_estimator}) but instead take the average value in each tomographic bin to avoid effects such as galaxy detection biases \citep{Kannawadi19}.}

We extract the cosmological information from the shear correlation signal using COSEBIs \citep[complete orthogonal sets of E/B-integrals,][]{Schneider10}. These present a method to cleanly decompose the shear 2PCFs into $E$- and $B$-modes by applying a set of oscillatory filter functions defined over a finite angular range between $\theta_{\rm min}$ and $\theta_{\rm max}$. The filter functions $T_{\pm n}(\theta)$ for the $n$-th COSEBI mode ($E_n$ and $B_n$) have exactly $n+1$ roots:
\begin{align}
    E_n &= \frac{1}{2} \int_{\theta_{\rm min}}^{\theta_{\rm max}} {\rm d}{\theta} \, \theta \left[ T_{+n}(\theta) \, \xi_+(\theta) + T_{-n}(\theta) \, \xi_-(\theta) \right] \\
    B_n &= \frac{1}{2} \int_{\theta_{\rm min}}^{\theta_{\rm max}} {\rm d}{\theta} \, \theta \left[ T_{+n}(\theta) \, \xi_+(\theta) - T_{-n}(\theta) \, \xi_-(\theta) \right]
\end{align}
One of the advantages of this formalism compared to the classical 2PCFs is that COSEBIs are less sensitive to small scales, where the complex physics of baryon feedback plays an important role, if a subset of the modes is chosen accordingly \citep{Asgari20}.

\subsection{Analysis pipeline}
\label{sec:pipe}

Our analysis pipeline is an upgraded version of {\tt CosmoPipe}\footnote{\url{https://github.com/AngusWright/CosmoPipe}} \citep{Wright20b} which is a wrapper for {\tt Cat\_to\_Obs}\footnote{\url{https://github.com/KiDS-WL/Cat\_to\_Obs\_K1000\_P1}} \citep{Giblin21} and the KiDS Cosmology Analysis Pipeline\footnote{\url{https://github.com/KiDS-WL/kcap}} (KCAP, \citealt{Joachimi21,Asgari21,Heymans21,Troester21}) that have both been used previously to analyse the KiDS-1000 data. The pipeline measures the shear 2PCFs using {\sc TreeCorr} \citep{Jarvis04,Jarvis15} on angular scales between \SI{0.5}{\arcmin} and \SI{300}{\arcmin} from which we compute the first five COSEBI modes using the logarithmic versions of the filter functions $T_{\pm n}(\theta)$. The logarithmic versions achieve a better compression of the cosmological signal onto fewer COSEBI modes.

We use the {\sc CosmoSIS} framework \citep{Zuntz15} to compute theoretical predictions with the KCAP COSEBI module \citep{Asgari12}. The linear matter power spectrum is modelled with CAMB \citep{Lewis00,Howlett12} and its non-linear evolution with {\sc HMCode} \citep{Mead15,Mead16}, whereas intrinsic alignments are calculated based on the model of \citet{Hirata04,Bridle07}. We then compare these predictions to the measured COSEBIs by sampling a Gaussian likelihood with {\sc MultiNest} \citep{Feroz09} using the analytical covariance model and priors of \citet{Joachimi21}. From this we infer constraints on the cosmological parameters of a spatially flat ΛCDM model. We additionally marginalise over a set of sample-dependent nuisance parameters which capture uncertainties in the shear and redshift calibration. Since Monte-Carlo samplers like {\sc MultiNest} are not designed to find the best fitting model parameters, we additionally run a Nelder-Mead minimiser \citep{Nelder65} starting from the maximum posterior point of all chains.

Based on this we quote parameter constraints and their uncertainty as the fit parameter value and the projected joint highest posterior density (PJ-HPD) that we obtain from the {\sc MultiNest} chains. We note that both best fit parameters as well as the PJ-HPD have statistical uncertainties of about 0.1σ or \SI{10}{\percent} on the 1σ constraints due to the limited number of posterior samples \citep{Joachimi21}.

\subsubsection{Redshift uncertainty}

We propagate uncertainties in the redshift calibration to the cosmological constraints by allowing the redshift distribution of each tomographic bin $i$ to vary by a shift $\delta z_i$. We use a set of correlated Gaussian priors $\delta z_i \sim N(\mu_i, \sigma_i)$ which allows us to apply an empirical redshift bias correction by choosing offsets $\mu_i \neq 0$. These offsets and their correlations (Table~\ref{tab:mbias}) are calibrated from spectroscopic and KiDS-like mock data \citep{vandenBusch20} in \citet{Hildebrandt21}. Since our analysis uses different calibration samples with altered sample selections, we would in principle need to perform a similar mock data analysis to recalibrate the priors for $\delta z_i$. However, these new samples contain many new spectroscopic datasets and the inclusion of photometric redshifts presents an additional challenge when attempting to model realistic photo-$z$ errors. Therefore, we assume that the variance of the KiDS-1000 priors is conservative enough to absorb potential changes of the redshift biases from KiDS-1000 to the new gold samples.

\subsubsection{Multiplicative shear uncertainty}

\begin{figure}
    \centering
    \includegraphics[width=\columnwidth]{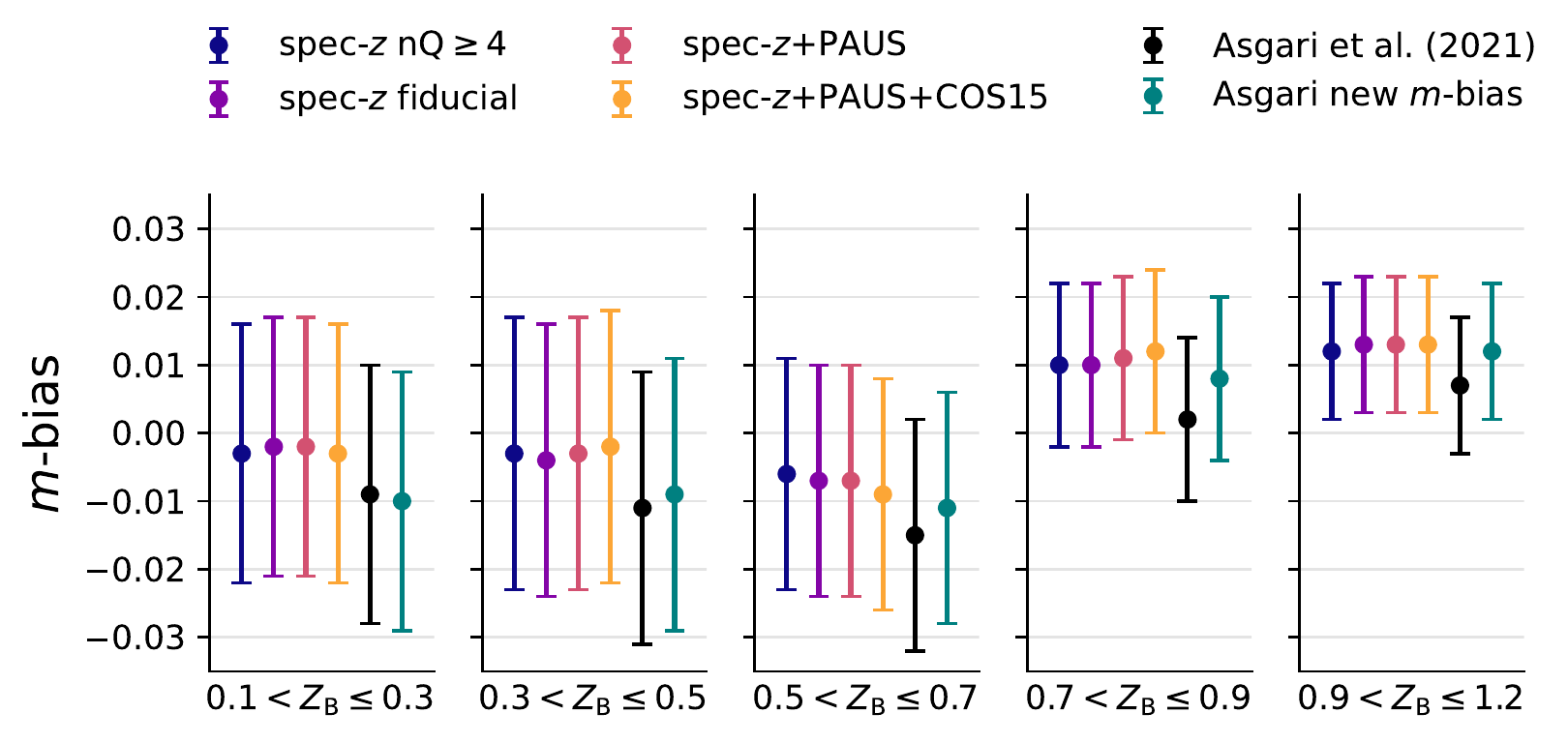}
    \caption{
        The $m$-bias values calculated for each gold sample per tomographic bin. Additionally, the values used by \citet{Asgari21} are shown in black and their revised values in teal.}
    \label{fig:mbias}
\end{figure}

The second set of sample-dependent nuisance parameters is the average multiplicative shear bias ($m$-bias, see Eq.~\ref{eq:shear_estimator}) in each tomographic bin. The effect of the $m$-bias and its uncertainty on the COSEBIs is captured in the covariance matrix and is calibrated by comparing the true galaxy ellipticities to those measured from a suite of image simulations generated by \citet{Kannawadi19}. The $m$-bias values vary little from sample to sample but are by up to 0.5σ larger than those of the KiDS-1000 sample \citep{Giblin21}. This led to the discovery of an issue with the way KiDS galaxies are assigned to galaxies in the COSMOS field which provides us with accurate shape information. We recompute the $m$-bias (Table~\ref{tab:mbias}, Fig.~\ref{fig:mbias}) and find that the revised values are in good agreement with those of the new gold samples. The updated values are also well within the uncertainty on $m$ that was accounted for in \citetalias{Asgari21}.

\begin{table}
    \centering
    \caption{
        The revised multiplicative shear bias $m_{\rm new}$ for the KiDS-1000 sample compared to the original values $m_{\rm old}$ of \citet{Asgari21}.}
    \renewcommand{\arraystretch}{1.10}
    \label{tab:mbias}
    \resizebox{\columnwidth}{!}{
    \begin{tabular}{cccccc}
        \hline\hline
        Bin & Photo-$z$ range & $m_{\rm old}$ & $m_{\rm new}$ & $\sigma_m$ & $\Delta z = z_{\rm est} - z_{\rm true}$ \\
        \hline
        1   & $0.1 < Z_{\rm B} \leq 0.3$ & $           -0.009$ & $           -0.010$ & $0.019$ & $\hphantom{-}0.000 \pm 0.0096$ \\
        2   & $0.3 < Z_{\rm B} \leq 0.5$ & $           -0.011$ & $           -0.009$ & $0.020$ & $\hphantom{-}0.002 \pm 0.0114$ \\
        3   & $0.5 < Z_{\rm B} \leq 0.7$ & $           -0.015$ & $           -0.011$ & $0.017$ & $\hphantom{-}0.013 \pm 0.0116$ \\
        4   & $0.7 < Z_{\rm B} \leq 0.9$ & $\hphantom{-}0.002$ & $\hphantom{-}0.008$ & $0.012$ & $\hphantom{-}0.011 \pm 0.0084$ \\
        5   & $0.9 < Z_{\rm B} \leq 1.2$ & $\hphantom{-}0.007$ & $\hphantom{-}0.012$ & $0.010$ & $-0.006 \pm 0.0097$ \\
        \hline
    \end{tabular}
    }
    \renewcommand{\arraystretch}{1.0}
    \tablefoot{
        We note that the uncertainties $\sigma_m$ remain unchanged. The last column summarises the parameters of the correlated Gaussian redshift priors ($\mu_i \pm \sigma_i$, where $\sigma_i$ is the square root of the diagonal element of the covariance matrix).}
\end{table}

\section{Results}
\label{sec:results}

In this section we present the new KiDS gold samples and cosmological constraints from an analysis of COSEBIs.

\subsection{New KiDS gold samples}
\label{sec:newgold}

Similar to the calibration samples (Sect.~\ref{sec:compilation}) we divide the gold samples in two categories: primary, which are based on the full compilation of spectroscopic data plus optionally photo-$z$ ({\sl \mbox{spec-$z$} fiducial}, {\sl \mbox{spec-$z$}+\allowbreak PAUS}, and {\sl \mbox{spec-$z$}+\allowbreak PAUS+\allowbreak COS15}, see Sect.~\ref{sec:comp_primary}), and secondary samples, which are calibrated with subsets of the spectroscopic calibration sample ({\sl \mbox{spec-$z$} \mbox{nQ $\geq 4$}}) or by using exclusively photo-$z$ ({\sl only-PAUS}, {\sl only-COS15}, and {\sl only-PAUS+\allowbreak COS15}, see Sect.~\ref{sec:comp_secondary}).

\subsubsection{Primary gold samples}
\label{sec:newgold_mix}

\begin{figure}
    \centering
    \includegraphics[width=0.95\columnwidth]{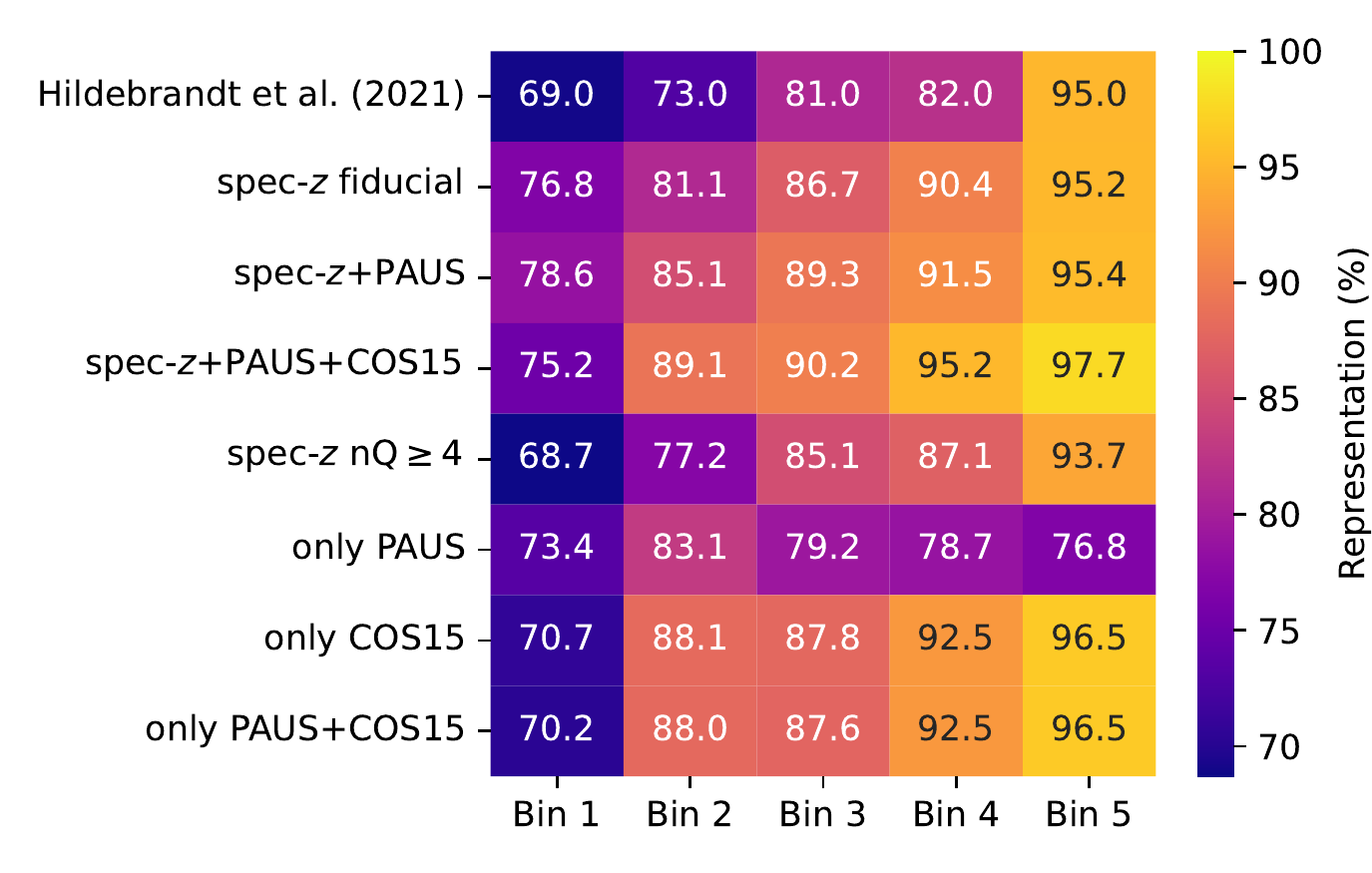}
    \caption{
        Representation fractions, the effective number density of the different gold samples relative to the full KiDS-1000 source sample, per tomographic bin. The effective number density factors in the lensing weight of each object and is calculated according to Eq.~(C.12) in \citet{Joachimi21}.}
    \label{fig:representation}
\end{figure}

We make a quantitative comparison of the selection of the three primary KiDS gold samples based on the representation fraction of each tomographic bin, the effective sample number density compared to the density of the full KiDS-1000 source sample. These are summarised for all gold samples in Fig.~\ref{fig:representation}. The numbers show that our new spectroscopic redshift compilation provides a much greater coverage of the KiDS source sample since our fiducial gold sample has a \SI{9}{\percent} higher accumulated number density than the previous data set calibrated by \citet{Hildebrandt21}, increasing the total representation fraction from \SI{80}{\percent} to \SI{89}{\percent}. In comparison to the former, our {\sl \mbox{spec-$z$} fiducial} gold sample and those constructed by the addition of the PAUS and COSMOS2015 photo-$z$ steadily increase the coverage fraction of KiDS galaxies across all tomographic bins, rising from \SI{73}{\percent}, \SI{81}{\percent}, and \SI{82}{\percent} to \SI{88}{\percent}, \SI{90}{\percent}, and \SI{95}{\percent} in bins 2, 3, and 4 respectively. The fifth tomographic bin, which contributes most of the cosmological signal in the cosmic shear analysis, shows the least change in its representation fraction due to the already very high coverage of \SI{95}{\percent} reported by \citet{Hildebrandt21}.

The only exception to the steadily increasing representation fractions is the first bin of {\sl \mbox{spec-$z$}+\allowbreak PAUS+\allowbreak COS15}, where the number density is about \SI{2.5}{\percent} lower compared to the fiducial case. This is the result of the quality control (Eq.~\ref{eq:qc}) removing some SOM groupings due to discrepancies in $\langle z_{\rm cal} \rangle$ and $\langle z_{\rm B} \rangle$, which arise when adding calibration sources and\slash or changing the gold selection. Expanding the calibration sample may shift $\langle z_{\rm cal} \rangle$ significantly, in particular in sparsely occupied SOM groupings, such that $\langle z_{\rm cal} \rangle - \langle z_{\rm B} \rangle$ exceeds the quality control threshold $5\sigma_{\rm mad}$, which will flag and exclude the corresponding KiDS galaxies from the gold sample.

\begin{figure*}
    \centering
    \includegraphics[width=0.75\textwidth]{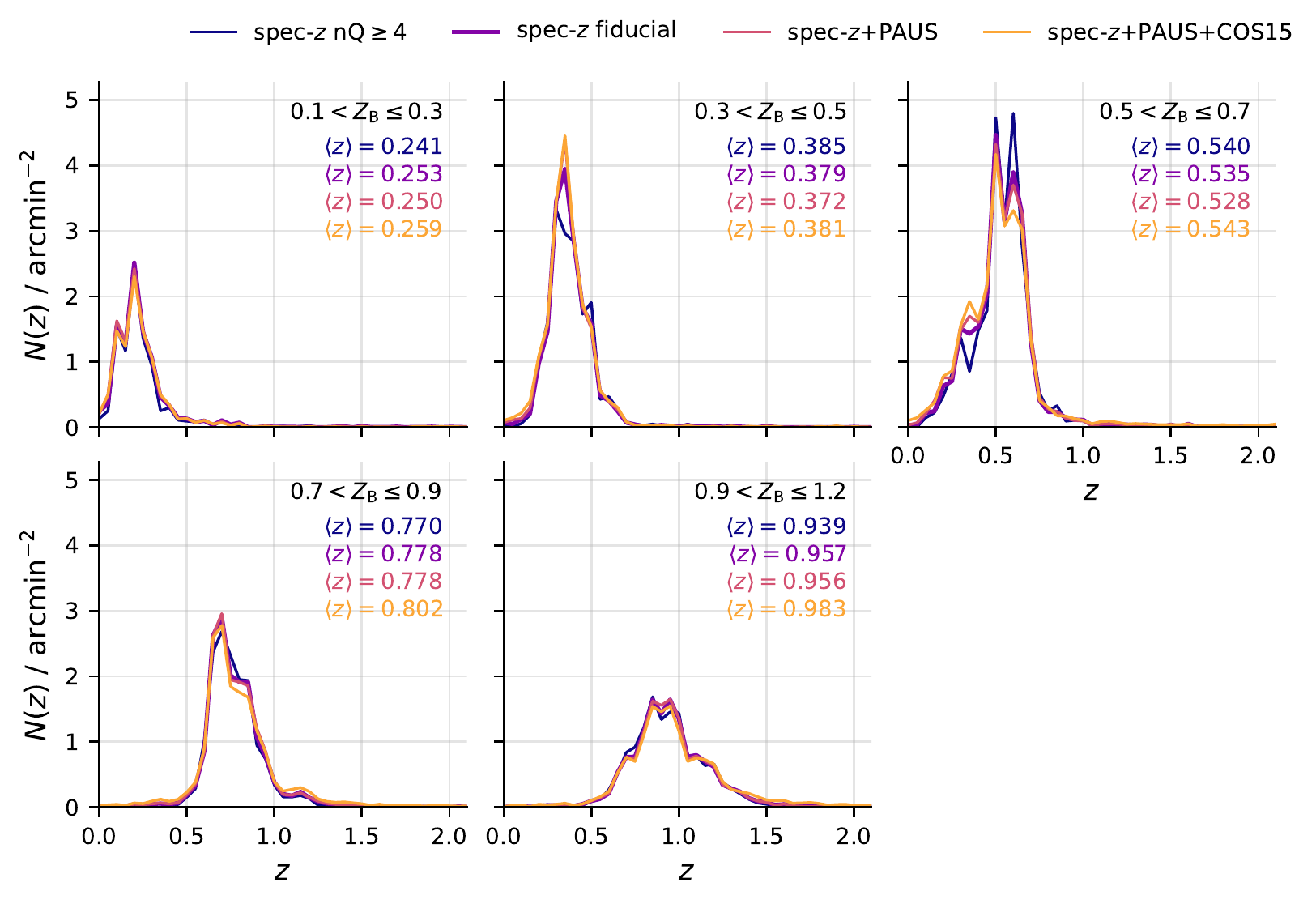}
    \caption{
        Comparison of the gold sample redshift distributions and their mean redshifts for all tomographic bins obtained from the different subsets of the calibration sample. We note that they all describe different subsets of the full KiDS-1000 source sample.}
    \label{fig:nz_compare}
\end{figure*}

The SOM redshift distributions of the gold samples are shown in Fig.~\ref{fig:nz_compare}; note that these samples do not represent the same galaxies. A comparison reveals two effects when adding photo-$z$ to the calibration sample: First, the bulk of the redshift distributions is skewed to lower redshifts as we add more data to the calibration sample which is most evident in the third tomographic bin. Secondly, COSMOS2015 adds a significant portion of high redshift objects to the compilation that extends the coverage of KiDS galaxies to higher redshifts, enhancing the tails of the redshift distributions and significantly increasing the mean redshifts, in particular of bin 5.

\begin{figure*}
    \centering
    \includegraphics[width=0.75\textwidth]{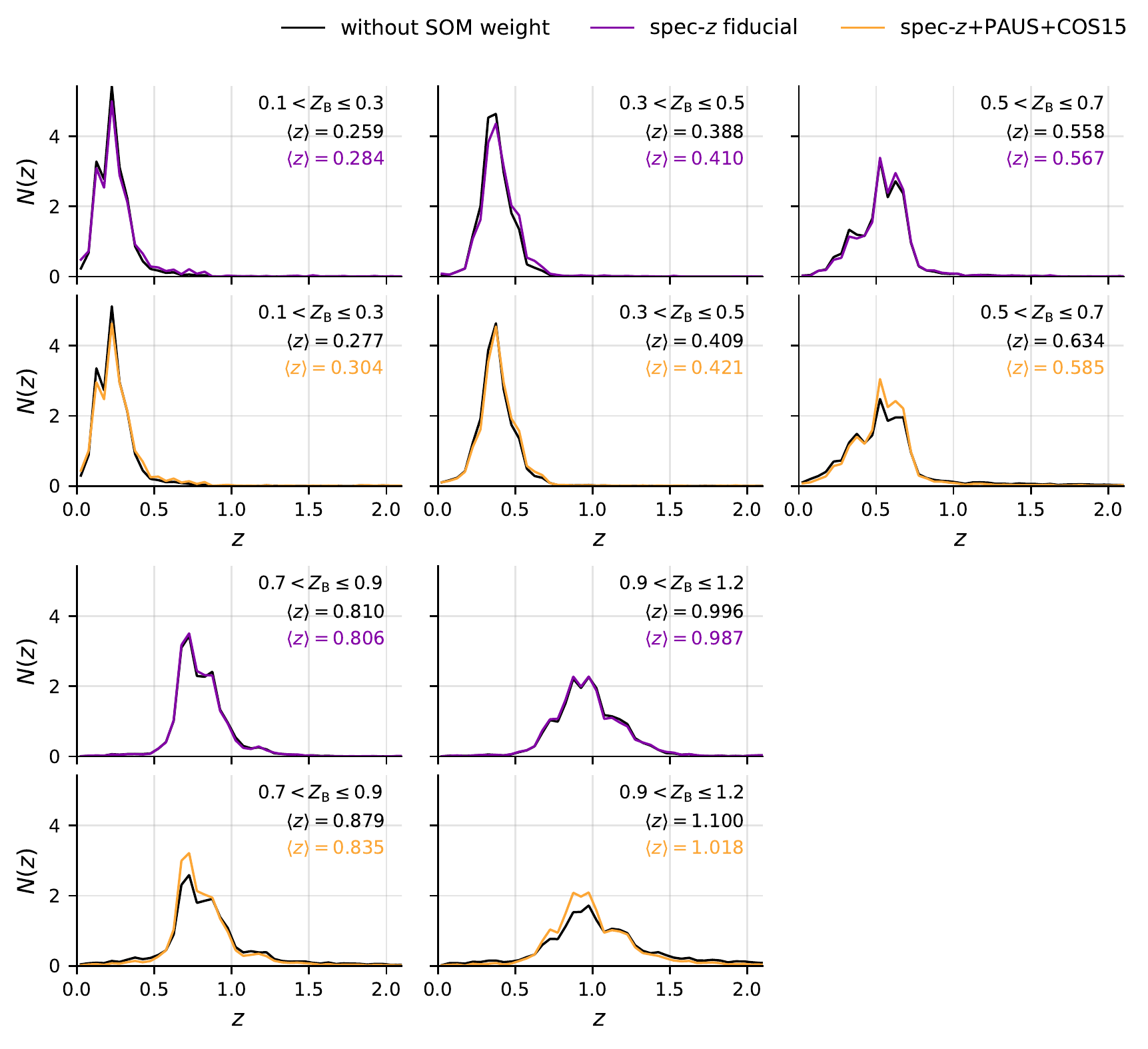}
    \caption{
        Comparison of the tomographically binned calibration sample (black lines) to the gold sample redshift distributions (coloured lines). The greater the difference between the black and the coloured lines, the more weighting is applied by the SOM to match the calibration sample to the KiDS data.}
    \label{fig:nz_somweight_groups}
\end{figure*}

Finally we compare the calibration sample redshift distributions in each tomographic bin to the resulting gold sample redshift distributions (Fig.~\ref{fig:nz_somweight_groups}). For most of the samples these distributions have very similar shapes and with mean redshifts agreeing within $\pm 0.02$, which indicates that the re-weighting (Eq.~\ref{eq:som_weight}) of the SOM groupings is very small on average. This changes once the COSMOS2015 redshifts are added to the calibration sample. Due to their significantly higher depth and mean redshifts, the redshift tails must be down-weighted significantly (up-weighted in bin 1 and 2) to match the density of the KiDS source sample. The down-weighting of the low redshift tails in the upper three bins can be explained by the fact that COSMOS2015 adds faint galaxies at these redshifts. The corresponding KiDS galaxies have a low lensing weight, which must be compensated by the SOM cell weights.

\subsubsection{Secondary gold samples}
\label{sec:newgold_phot}

The first of our secondary gold samples we calibrate using only those galaxies of the spectroscopic compilation that have the most secure spectroscopic redshifts of at least \SI{99}{\percent} confidence. Due to similarities to the SOM calibration sample used by \citet[see Sect.~\ref{sec:compilation}]{Hildebrandt21}, this {\sl \mbox{spec-$z$} \mbox{nQ $\geq 4$}} gold sample positions itself in between the latter and the {\sl \mbox{spec-$z$} fiducial} sample in terms of representation fractions. In tomographic bin 4 and 5 however it is lacking some of the high redshift sources due to the more conservative spectroscopic flagging, reducing both the mean redshifts (Fig.~\ref{fig:nz_compare}) as well as the representation fractions (Fig.~\ref{fig:representation}).

The remaining three gold samples exclusively rely on photometric redshifts from PAUS and COSMOS2015 (Sect.~\ref{sec:comp_secondary}). Due to the great overlap of sources, both samples that contain the COSMOS2015 photo-$z$ ({\sl only-COS15} and {\sl only-PAUS+\allowbreak COS15}) achieve representation fractions which are just \SI{2}{\percent} smaller than those of the full redshift compilation, again due to the quality control (Eq.~\ref{eq:qc}). As seen for {\sl \mbox{spec-$z$}+\allowbreak PAUS+\allowbreak COS15}, the effective density in the first bin is significantly reduced. The {\sl only-PAUS} gold sample exhibits significantly suppressed tails in bin 4 and 5 due to a lack of high redshift sources that only the much deeper COSMOS2015 provides. This also imprints on the representation fractions which are similar to the fiducial gold sample in bin 1 and 2 but are about \SI{10}{\percent} lower in bin 3 and 4 and \SI{18}{\percent} lower in the fifth tomographic bin. This example highlights the abilities of the SOM to flag and remove sources that cannot be calibrated by the DIR approach with the particular calibration sample.

\subsection{Cosmological constraints}
\label{sec:constraints}

\begin{figure}
    \centering
    \includegraphics[width=\columnwidth]{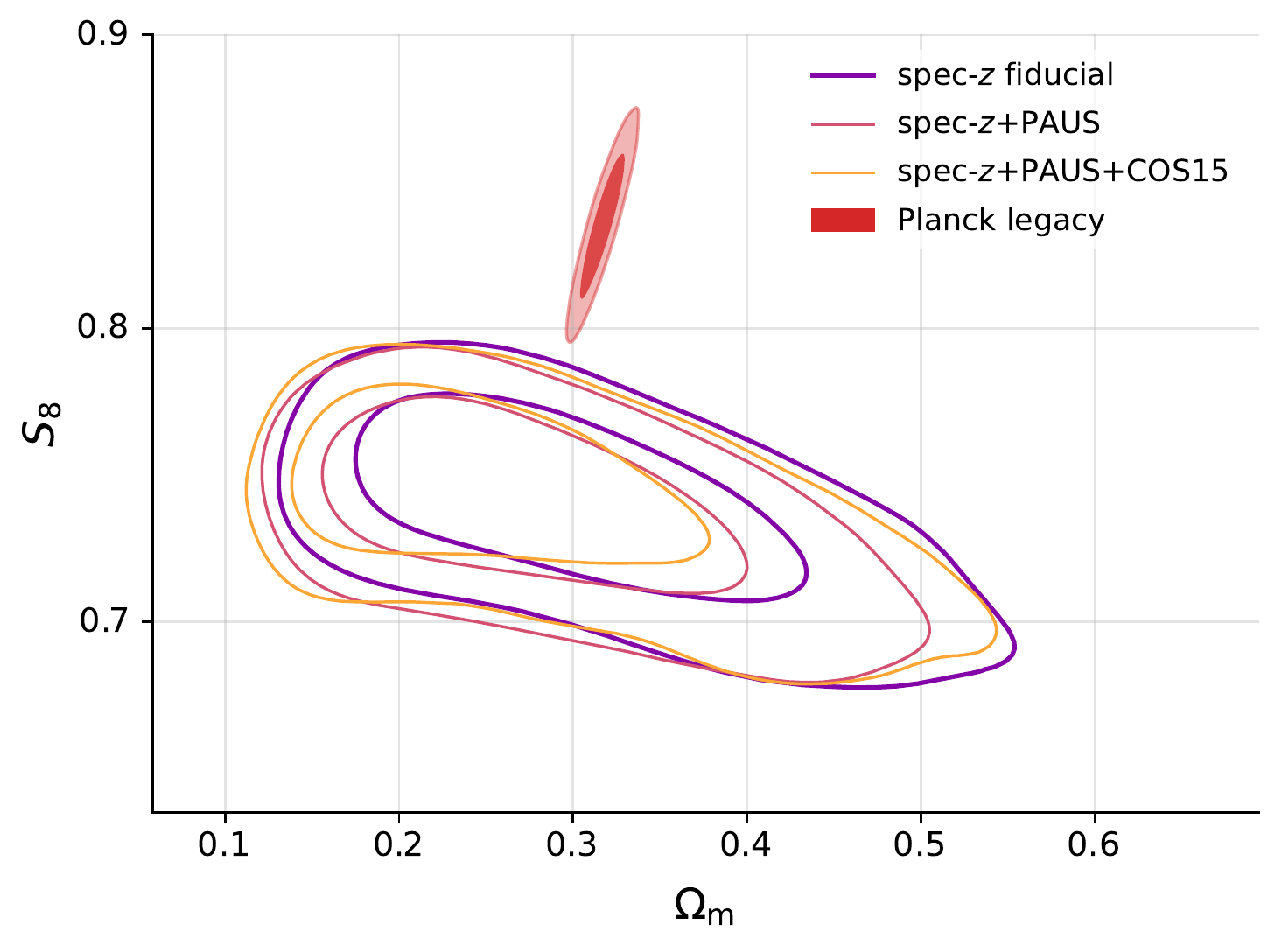}
    \caption{
        Marginalised constraints for the joint distributions of $S_8$ and $\Omega_{\rm m}$ (\SI{68}{\percent} and \SI{95}{\percent} credible regions) obtained for different gold samples and Planck legacy (TT, TE, EE + lowE). Since the contours represent different galaxy samples, some deviation is expected.}
    \label{fig:S8_Om}
\end{figure}

\begin{figure}
    \centering
    \includegraphics[width=\columnwidth]{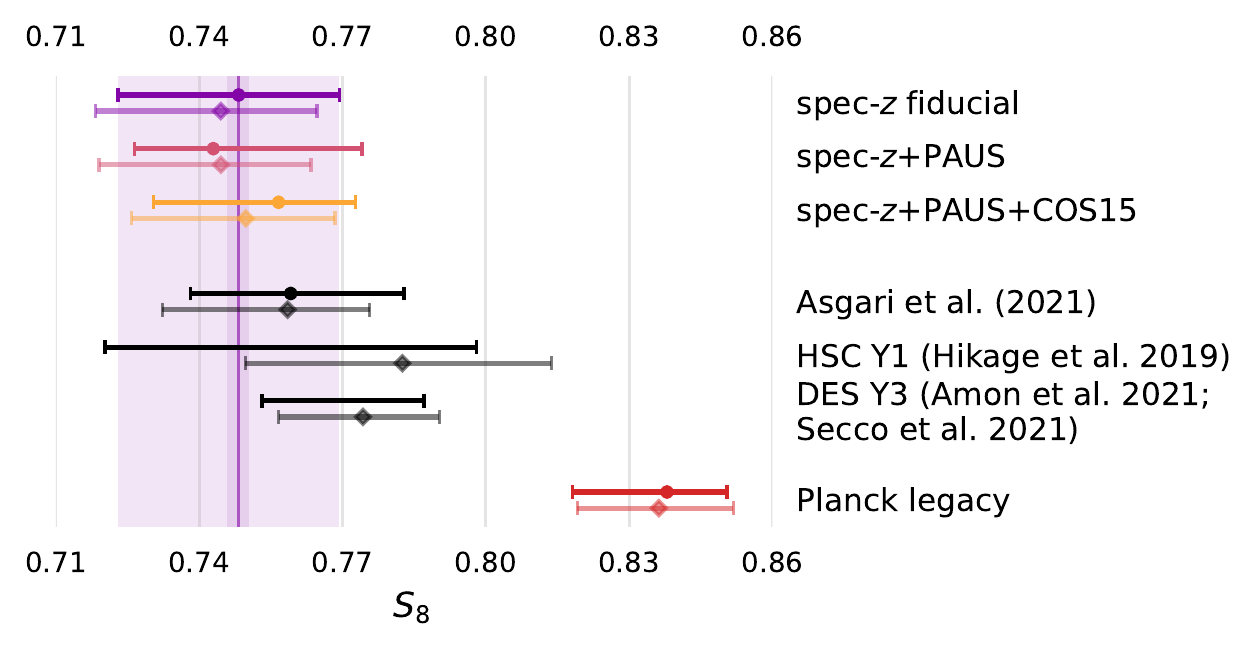}
    \caption{
        Comparison of the $S_8$ constraints from our gold samples to other studies. We show the best fit (where available) and 68th-percentile PJ-HPD (circles, opaque data points) and the maximum of the marginal distribution and the associated 68th-percentile (diamonds, semi-transparent). We compare to \citet{Asgari21}, HSC-Y1 \citep{Hikage19}, DES-Y3 \citep{Amon21}, and Planck legacy. The coloured vertical line and outer bands indicate the constraints from the fiducial gold sample, the inner bands the expected variance of the sampler.}
    \label{fig:S8}
\end{figure}

We present cosmological results for our primary KiDS gold samples, focusing on a relative comparison to \citetalias{Asgari21} and other literature values. We summarise the numerical values of the most relevant cosmological parameters in Table~\ref{tab:params}. In Fig.~\ref{fig:S8_Om} and~\ref{fig:S8} we highlight comparisons of the derived parameter $S_8 = \sigma_8 (\Omega_{\rm m} / 0.3)^{0.5}$, which is the primary measurable of weak lensing due to the degeneracy between $\Omega_{\rm m}$, the dimensionless matter density parameter, and $\sigma_8$, parameterising the amplitude of the linear power spectrum.

\begin{table*}
    \centering
    \caption{
        Summary of the main cosmological parameter constraints (best fit and 68th-percentile PJ-HPD) from COSEBIs for all gold samples and their comparison to \citet{Asgari21} and Planck legacy (TT, TE, EE + lowE).}
    \renewcommand{\arraystretch}{1.33}
    \label{tab:params}
    \begin{tabular}{lcccccc}
\hline\hline
Sample & $\chi^2$ &                           $A_{\rm IA}$ &           $\Omega_{\rm m}$ &                 $\sigma_8$ &                      $S_8$ &                 $\Sigma_8$ \\
\hline
spec-$z$ fiducial             &     63.2 &  $\hphantom{-}0.301_{-0.377}^{+0.343}$ &  $0.268_{-0.055}^{+0.126}$ &  $0.791_{-0.155}^{+0.110}$ &  $0.748_{-0.025}^{+0.021}$ &  $0.744_{-0.021}^{+0.017}$ \\
spec-$z$+PAUS                 &     64.0 &             $-0.097_{-0.416}^{+0.422}$ &  $0.226_{-0.053}^{+0.108}$ &  $0.857_{-0.164}^{+0.130}$ &  $0.743_{-0.016}^{+0.031}$ &  $0.732_{-0.012}^{+0.028}$ \\
spec-$z$+PAUS+COS15           &     68.3 &  $\hphantom{-}0.164_{-0.428}^{+0.450}$ &  $0.194_{-0.048}^{+0.109}$ &  $0.940_{-0.200}^{+0.144}$ &  $0.757_{-0.026}^{+0.016}$ &  $0.740_{-0.025}^{+0.013}$ \\
spec-$z$ $\mathrm{nQ} \geq 4$ &     61.6 &             $-0.033_{-0.393}^{+0.408}$ &  $0.195_{-0.046}^{+0.079}$ &  $0.951_{-0.171}^{+0.113}$ &  $0.766_{-0.023}^{+0.023}$ &  $0.750_{-0.019}^{+0.021}$ \\
only PAUS                     &     74.1 &             $-0.186_{-0.345}^{+0.526}$ &  $0.203_{-0.053}^{+0.108}$ &  $0.916_{-0.183}^{+0.163}$ &  $0.752_{-0.019}^{+0.028}$ &  $0.737_{-0.017}^{+0.023}$ \\
only COS15                    &     67.2 &  $\hphantom{-}0.164_{-0.545}^{+0.339}$ &  $0.253_{-0.074}^{+0.132}$ &  $0.814_{-0.177}^{+0.157}$ &  $0.747_{-0.033}^{+0.013}$ &  $0.741_{-0.023}^{+0.017}$ \\
only PAUS+COS15               &     67.1 &  $\hphantom{-}0.078_{-0.369}^{+0.515}$ &  $0.216_{-0.033}^{+0.176}$ &  $0.876_{-0.243}^{+0.085}$ &  $0.744_{-0.028}^{+0.020}$ &  $0.731_{-0.017}^{+0.024}$ \\
Asgari et al. (2021)          &     82.2 &  $\hphantom{-}0.264_{-0.336}^{+0.424}$ &  $0.246_{-0.060}^{+0.101}$ &  $0.838_{-0.141}^{+0.140}$ &  $0.759_{-0.021}^{+0.024}$ &  $0.751_{-0.016}^{+0.024}$ \\
Planck legacy                 &      --- &                                    --- &  $0.319_{-0.010}^{+0.006}$ &  $0.813_{-0.008}^{+0.007}$ &  $0.838_{-0.020}^{+0.013}$ &  $0.841_{-0.021}^{+0.013}$ \\
\hline
\end{tabular}

    \renewcommand{\arraystretch}{1.0}
    \tablefoot{
        Additionally shown are the $\chi^2$ values (for $70.5$ effective degrees of freedom) and $\Sigma_8 = \sigma_8 (\Omega_{\rm m} / 0.3)^\alpha$ calculated for \protect$\alpha_{\rm fid} = 0.55$%
, determined from a fit to the posterior samples of the fiducial chain.}
\end{table*}

First we ensure that our wrapper for the cosmological pipeline delivers results that are consistent with those of \citetalias{Asgari21}. We re-analyse the original KiDS-1000 data, running our pipeline with the same input parameters (`Asgari reanalysis'). We find that the constraints on all cosmological parameters agree within the expected variance of the Monte-Carlo sampler. The best fit solution has a slightly larger $\chi^2$ of $83.8$ compared to $82.2$ for KiDS-1000 which is driven primarily by the slight change in the data vector (see App.~\ref{app:K1000diff}). Similarly, the correction of the $m$-bias values (Table~\ref{tab:mbias}) has no significant impact on the cosmological constraints either (Table.~\ref{tab:params_A21}).

\subsubsection{Primary gold samples}

For our KiDS gold samples we find a tendency to lower $S_8$-values, albeit at low significance (Figs.~\ref{fig:S8_Om}, \ref{fig:S8} and Table~\ref{tab:params}). In particular we obtain $S_8 = 0.748_{-0.025}^{+0.021}$%
 for the {\sl \mbox{spec-$z$} fiducial} and $S_8 = 0.743_{-0.016}^{+0.031}$%
 for the {\sl \mbox{spec-$z$}+\allowbreak PAUS} gold samples which is about 0.4σ lower compared to \citetalias{Asgari21} and exceeds the expected statistical variance of the sampler (about 0.1σ). Deviations to some degree between these gold samples are expected, since they all represent different galaxy selections. The goodness of fit improves for all gold samples but in particular for {\sl \mbox{spec-$z$} fiducial}, where it reduces to $\chi^2 \approx 63$ compared to 82 in KiDS-1000. The {\sl \mbox{spec-$z$}+\allowbreak PAUS+\allowbreak COS15} sample on the other hand is in very good agreement with the original KiDS-1000 analysis.

In addition to $S_8$ we consider the more general $\Omega_{\rm m}$-$\sigma_8$-degeneracy case of $\Sigma_8 = \sigma_8 (\Omega_{\rm m} / 0.3)^\alpha$, where $\alpha$ is a free parameter. We estimate the reference value for $\alpha$ by fitting the posterior samples of the fiducial chain and obtain $\alpha_{\rm fid} = 0.55$%
 which is close to $\alpha=0.54$ for COSEBIs in \citetalias{Asgari21}. This projection optimises the signal-to-noise-ratio compared to $S_8$ and we find $\Sigma_8 = 0.744_{-0.021}^{+0.017}$ for {\sl \mbox{spec-$z$} fiducial} compared to $\Sigma_8 = 0.751_{-0.016}^{+0.024}$ which we obtain for \citetalias{Asgari21} with $\alpha=\alpha_{\rm fid}$. Furthermore, the scatter of $\Sigma_8$ for the different gold samples is significantly smaller than the scatter in $S_8$ (see Table~\ref{tab:params}).

If we compare the marginal errors of $S_8$ for the different gold samples (Table~\ref{tab:params_marginal}), which have a smaller statistical variance than the PJ-HPD, we find that the constraints improve by \SI{5}{\percent} when including PAUS and another \SI{4}{\percent} when including COSMOS2015. Since the constraints on $\Sigma_8$ are almost constant, these changes in $S_8$ are most probably related to small changes in the $\Omega_{\rm m}$-$\sigma_8$-degeneracy. We also compare $A_{\rm IA}$, which is the dimensionless amplitude of the intrinsic alignment galaxy power spectrum, and find that its value is stable within the uncertainties in all our analyses.


\section{Discussion}
\label{sec:discussion}

\subsection{Gold sample selection and calibration}
\label{sec:discuss_gold}

A side-by-side comparison of the different KiDS gold samples presented in Sect.~\ref{sec:newgold} is non-trivial. On the one hand adding or removing galaxies from the calibration sample changes the redshift distribution of each of the SOM groupings that, according to Eq.~(\ref{eq:som_weight}), determine the sample's redshift distribution. On the other hand two distinct gold samples are comprised of different galaxies since a modification of the calibration sample will also apply an implicit selection on the set of representative SOM groupings. Both of these effects combined determine the overall calibrated redshift distribution.

\begin{figure*}
    \centering
    \includegraphics[width=0.75\textwidth]{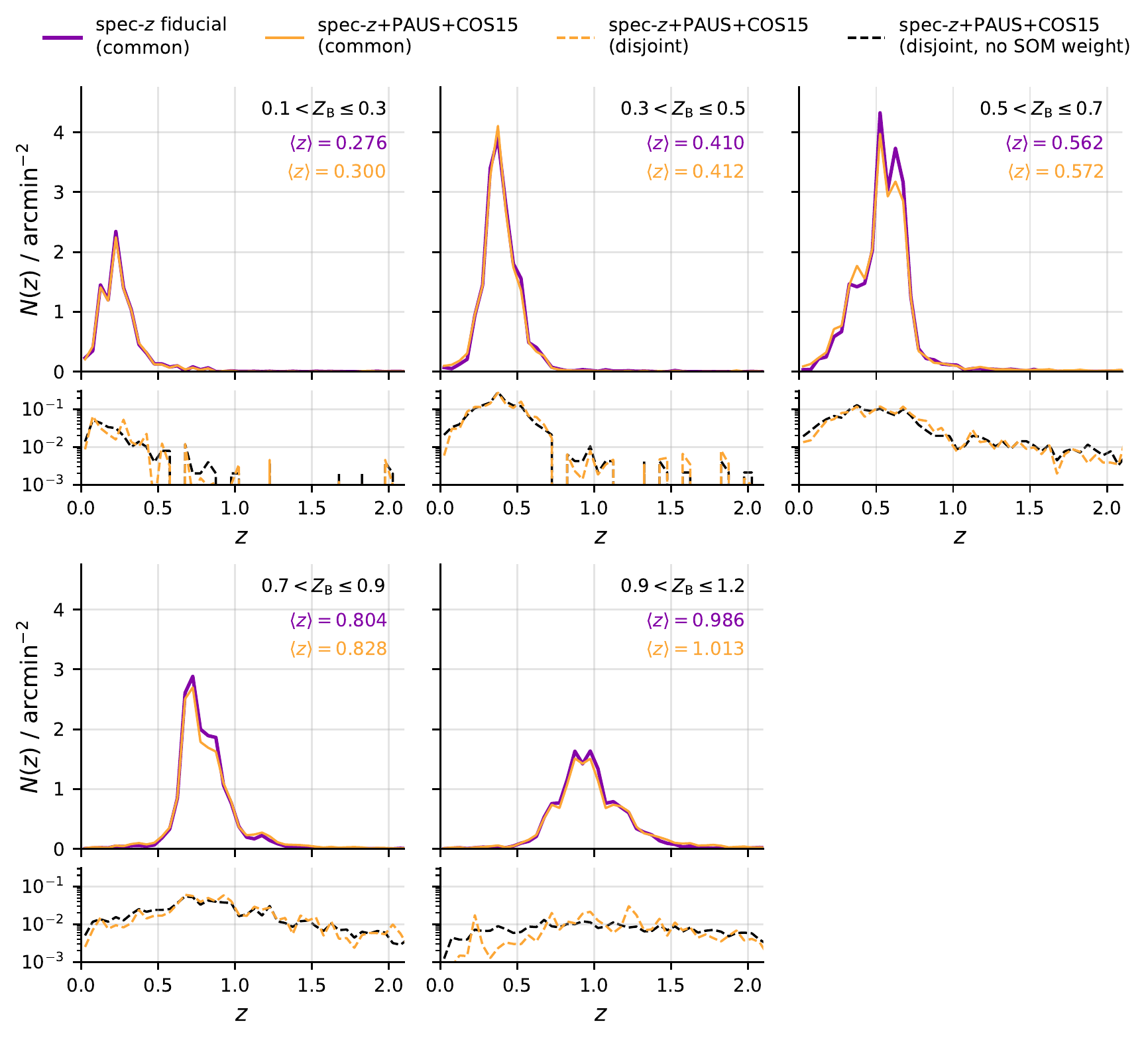}
    \caption{
        Comparison of the redshift distributions between the fiducial sample and the sample including the PAUS and COSMOS2015 photo-$z$. Top: The SOM redshifts derived from the subset of SOM groupings present in both samples. Bottom: The SOM redshift distributions of the groupings that are found only in the {\sl \mbox{spec-$z$}+\allowbreak PAUS+\allowbreak COS15} sample (dashed yellow) and the underlying redshift distribution in the calibration sample of the same groupings, i.e. not applying the SOM weighting in Eq.~(\ref{eq:som_weight}), scaled to the amplitude of the former for comparison.}
    \label{fig:nz_compare_groups}
\end{figure*}

This is exemplified by the fact that, as we expand the calibration sample from the {\sl \mbox{spec-$z$} \mbox{nQ $\geq 4$}} subset to {\sl \mbox{spec-$z$}+\allowbreak PAUS+\allowbreak COS15}, we generally see that both the representation fractions and the mean redshifts increase across all bins. The galaxies added in each iteration are typically fainter (with the exception of PAUS) at the cost of lower redshift accuracy, which allows us to calibrate additional KiDS galaxies, preferentially at the tails of the redshift distributions. At the same time we can observe in Fig.~\ref{fig:nz_compare} that the redshift distributions are skewed to lower redshifts, which can be explained by the fact that there are disproportionately many more galaxies with $z < 0.5$ added in each iteration to the calibration sample (Fig.~\ref{fig:nz_calib}). These in turn increase the representation fraction of low and intermediate redshift galaxies in the gold sample (see also Fig.~\ref{fig:representation}). This implies that we are changing the redshift calibration since the skewing applies to each individual SOM group. We can separate these two effects by splitting the {\sl \mbox{spec-$z$} fiducial} and {\sl \mbox{spec-$z$}+\allowbreak PAUS+\allowbreak COS15} gold samples in two subsets, one containing those SOM groupings that are common to both samples (i.e. containing the same KiDS galaxies) and groupings that can only be calibrated using the full redshift compilation. The subset of KiDS galaxies that is common to both gold samples shows the same redshift skewing as seen with all galaxies (top panels of Fig.~\ref{fig:nz_compare_groups}), whereas the additional COSMOS2015 galaxies contribute significantly at the low and high redshift tails of the tomographic bins (bottom panels of Fig.~\ref{fig:nz_compare_groups}).

The final ingredient to the redshift calibration is the quality control cut (Eq.~\ref{eq:qc}) that we apply to remove potentially miscalibrated parts of the colour space. This becomes most obvious when comparing the representation fractions of the first tomographic bin (Fig.~\ref{fig:representation}) which decrease whenever we add the COSMOS2015 data (compare {\sl \mbox{spec-$z$}+\allowbreak PAUS} to {\sl \mbox{spec-$z$}+\allowbreak PAUS+\allowbreak COS15} and {\sl only-PAUS} to {\sl only-PAUS+\allowbreak COS15}). The low redshift of the KiDS galaxies in this bin makes the sample particularly susceptible to the addition of high redshift galaxies. These can significantly change the mean redshift of the calibration sample $\langle z_{\rm cal} \rangle$ compared to the KiDS photo-$z$ in the SOM groupings which then may fail to pass the quality control (see also App.~\ref{app:SOMqc}). The great depth of the COSMOS2015 data compared to the spectroscopic data also explains why the SOM needs to apply more weighting to match the {\sl \mbox{spec-$z$}+\allowbreak PAUS+\allowbreak COS15} compilation to the KiDS colour-space (Fig~\ref{fig:nz_somweight_groups}).

\subsection{Cosmological constraints}

The gold sample selection effects are, due to their redshift dependence, directly propagated to the cosmological constraints (Sect.~\ref{sec:constraints}), causing shifts in $S_8$ of up to 0.5σ from sample to sample. One of the assumptions in our analysis is that we can adopt the same Gaussian priors for the $\delta z_i$ nuisance parameters (Sect.~\ref{sec:pipe}) that are used by \citetalias{Asgari21}. We therefore re-analyse the fiducial gold sample assuming no knowledge of the empirical redshift bias by centring the priors on $\mu_i = 0$. For this run we find that the value of $S_8 = 0.747_{-0.022}^{+0.022}$%
 is in good agreement with the fiducial analysis.

On the other hand it may seem that our choice for the widths $\sigma_i$ of the $\delta z_i$ priors may be insufficient to accommodate for the apparent variance in the mean redshifts of the different gold samples (see Fig.~\ref{fig:nz_compare}). This variance, however, is not only determined by potential systematic biases in the redshift calibration between any of these samples, but also by changes in the gold sample selection itself, as discussed above. Therefore, the question of the correct redshift prior can only be answered with realistic simulated data sets that are currently not available for our extended redshift calibration sample. Nevertheless, a comparison of the $S_8$ values allows us to get an estimate of the variance induced by the selection effects in the calibration data and the resulting parameter constraints from the gold samples.

\begin{figure*}
    \centering
    \includegraphics[width=\columnwidth]{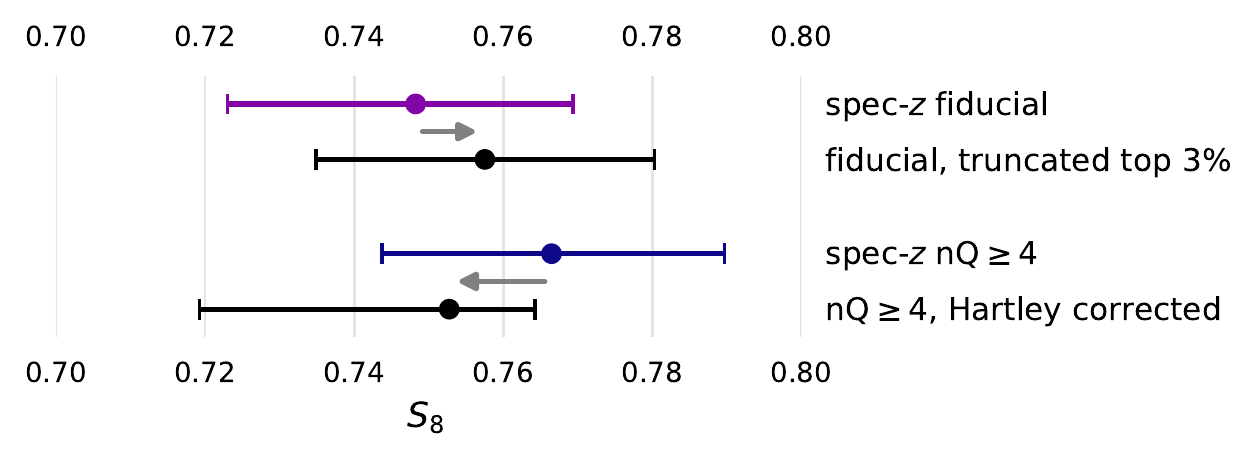}
    \hfill
    \includegraphics[width=\columnwidth]{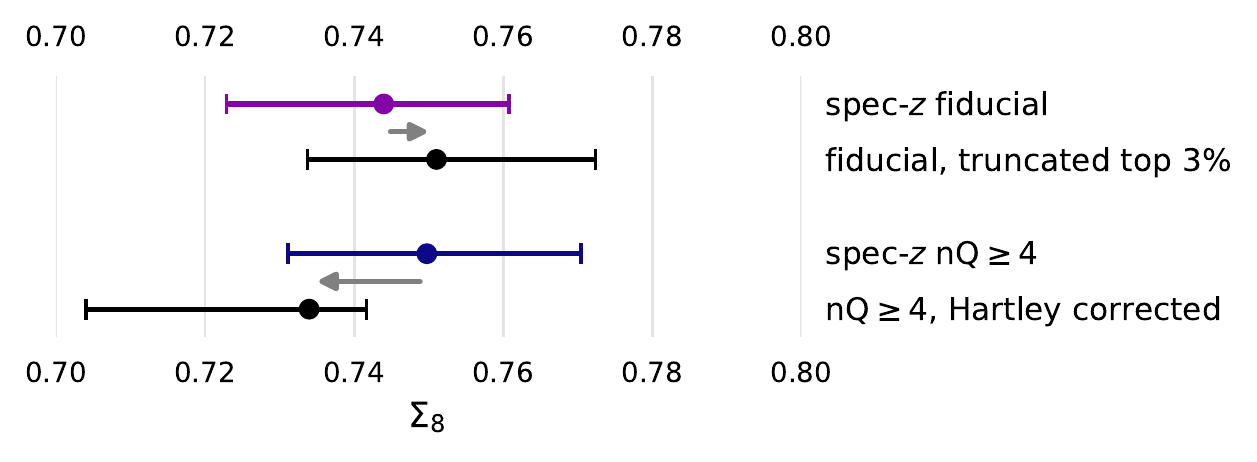}
    \caption{
        Comparison between constraints (best fit and 68th-percentile PJ-HPD) on $S_8$ (left) and $\Sigma_8$ (right) for {\sl \mbox{spec-$z$} fiducial}, {\sl \mbox{spec-$z$} \mbox{nQ $\geq 4$}}, and the corresponding test scenario A and B. The grey arrows indicate shifts in $S_8$ introduced in the tests by modifying the redshift distributions.}
    \label{fig:S8_sys}
\end{figure*}

\subsubsection{The fiducial gold sample}

Next to the {\sl \mbox{spec-$z$} fiducial} sample that is based on the full spectroscopic compilation we also define the {\sl \mbox{spec-$z$} \mbox{nQ $\geq 4$}} sample that relies only on the most secure redshifts. The estimated $S_8 = 0.766_{-0.023}^{+0.023}$%
 for the latter, arising from particularly low $\Omega_{\rm m}$ and high $\sigma_8$ values, is about 0.5σ larger than in the fiducial case and surpasses all other gold samples (Table~\ref{tab:params}). In the $\Sigma_8$ projection this difference reduces to 0.2σ.

The shift in $S_8$ between gold samples that are both calibrated with spectroscopic data begs the question which of these estimates is more reliable. The primary difference between the two calibration datasets is the selection using redshift quality flags. Selecting spectra based on the redshift confidence is a trade-off between constructing a sample that is confined to regions of the colour-redshift-space in which galaxies have distinct spectral features that allow secure redshift determination and a sample with an increasing fraction of galaxies with catastrophically misidentified redshifts. In the latter case the redshifts of the calibration sample themselves cause a biasing of the gold sample redshifts and in turn $S_8$. In case of selecting only the highest quality redshifts, the biases arises from a misrepresentation of the imaging data by the calibration sample, as shown by \citet{Hartley20}. The redshift distribution of the calibration sample in each SOM grouping depends on the quality flag and thus the relative representation of different galaxy populations in the gold sample may change. We investigate the magnitude of both these effects by assuming two worst-case scenarios which shift down the $S_8$ estimate obtained for {\sl \mbox{spec-$z$} fiducial} and shift up $S_8$ for {\sl \mbox{spec-$z$} \mbox{nQ $\geq 4$}}:

\begin{enumerate}[label=\Alph*)]
    \item We assume in case of the {\sl \mbox{spec-$z$} fiducial} sample that \SI{5}{\percent} of truly low redshift galaxies with $3 \leq {\rm nQ} < 4$ (nominal \SI{95}{\percent} certainty), as well as \SI{1}{\percent} in case of ${\rm nQ} \geq 4$ (\SI{99}{\percent} certainty), are catastrophically misidentified as high redshift galaxies. Since both redshift flags are equally common in the fiducial sample we expect a combined spectroscopic failure rate of about \SI{3}{\percent}. We implement this worst-case scenario on {\sl \mbox{spec-$z$} fiducial} by truncating the top \SI{3}{\percent} of all redshift distributions which should increase the recovered $S_8$ value. We calculate the redshift $z_{97}$ corresponding to the 97-th percentile of the redshift distribution $n(z)$, set $n(z)=0$ at $z > z_{97}$ and re-normalise to reproduce the original gold sample number density.
    \item We speculate that the calibration of the {\sl \mbox{spec-$z$} \mbox{nQ $\geq 4$}} sample suffers from the same spectroscopic misrepresentation effects studied by \citet[from Fig.~6 therein]{Hartley20}, who found redshift biases $\langle z \rangle - \langle z_{\rm true} \rangle$ of $0.008$, $0.022$, $-0.003$, and $-0.058$ in the four tomographic redshifts bins of simulated DES and spectroscopic data, for the first time implementing a realistic, simulated ${\rm nQ} \geq 4$ sample selection. Since we currently do not have comparable spectroscopic mock data in KiDS, we assume in this scenario that the bias applies at the same magnitude to the {\sl \mbox{spec-$z$} \mbox{nQ $\geq 4$}} gold sample. We therefore correct the assumed bias by interpolating the values from the four DES bins to the five tomographic bins of KiDS and shift the {\sl \mbox{spec-$z$} \mbox{nQ $\geq 4$}} redshift distributions by $-0.008$, $-0.015$, $-0.014$, $0.003$, and $0.058$. We consider this to be an even more conservative assumption than scenario A, since the bias should be significantly smaller in case of KiDS thanks to the 9-band imaging and the improvements of the SOM calibration over the classical DIR approach (Sect.~\ref{sec:SOM}) that is used by \citet{Hartley20}.
\end{enumerate}

With these modifications to the redshift distributions, scenario A ({\sl fiducial, truncated top \SI{3}{\percent}}) yields a higher and scenario B ({\sl\mbox{nQ $\geq 4$}, Hartley corrected}) a lower estimate for $S_8$ (Fig.~\ref{fig:S8_sys}, left). These results indicate that the combination of these two effects may explain the observed differences between $S_8$ in {\sl \mbox{spec-$z$} fiducial} and {\sl \mbox{spec-$z$} \mbox{nQ $\geq 4$}}. When comparing the projection $\Sigma_8$ instead (Fig.~\ref{fig:S8_sys}, right), which is less susceptible to shifts along the $\Omega_{\rm m}$-$\sigma_8$-degeneracy, the difference between {\sl \mbox{spec-$z$} fiducial} and {\sl \mbox{spec-$z$} \mbox{nQ $\geq 4$}} is much smaller. However the shift, introduced when correcting the redshift distributions in scenario B, is about twice as big compared to the $S_8$ case, indicating that the selection effects studied in \citet{Hartley20} can have a significant impact on cosmological constraints. Furthermore the completeness, which determines the ability to correctly map out colour-redshift degeneracies in each SOM grouping, and the quality of the calibration sample redshifts should be balanced carefully to minimise biases in the redshift calibration and the cosmological analysis.

\subsubsection{Other gold samples}

\begin{figure}
    \centering
    \includegraphics[width=\columnwidth]{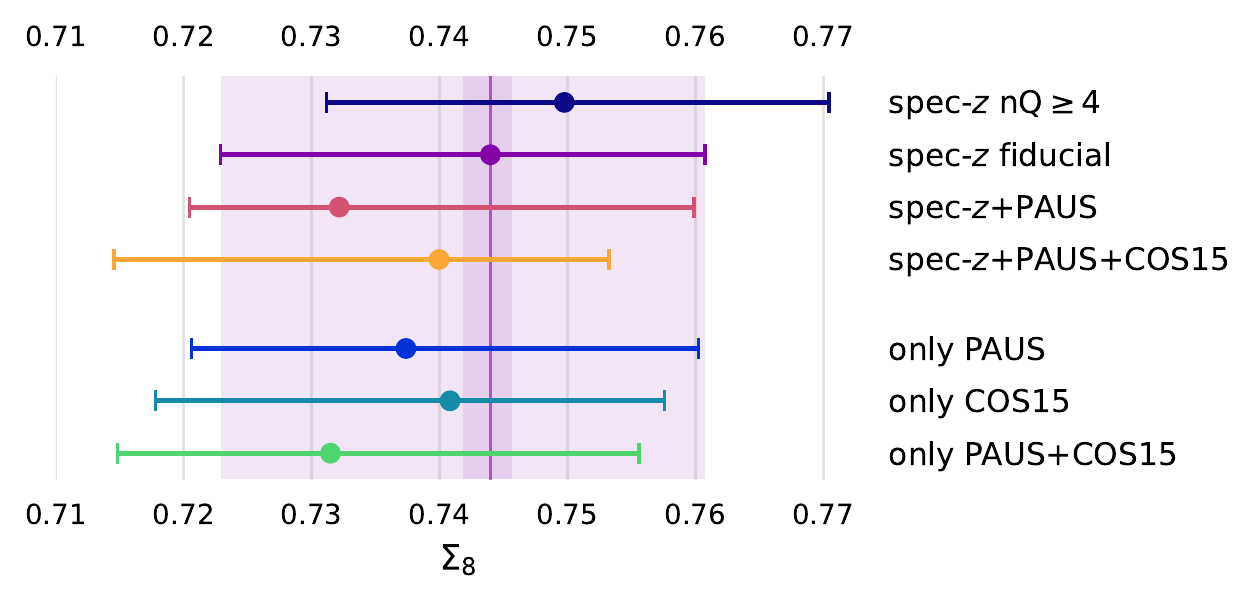}
    \caption{
        Constraints on $\Sigma_8$ (best fit and 68th-percentile PJ-HPD) for all primary and secondary gold samples. The coloured vertical line and outer bands indicate the constraints from the fiducial gold sample, the inner bands the expected variance of the sampler.}
    \label{fig:S8_tests}
\end{figure}

Finally, we make a relative comparison of the $\Sigma_8$ constraints from the remaining gold samples, since $\Sigma_8$ typically exhibits a smaller scatter than the corresponding $S_8$ values. The gold samples that are calibrated using only photo-$z$ from PAUS, COSMOS2015, or a combination of both, prefer smaller $\Sigma_8$ values compared to {\sl \mbox{spec-$z$} fiducial} (Fig.~\ref{fig:S8_tests}). While the cosmological constraints in a similar comparison between redshifts calibrated using spectroscopic data and COSMOS2015 \citep{Hildebrandt20} show a more pronounced shift in the opposite direction, this difference is caused by the gold sample selection \citep{Wright20b}. In our analysis the shifts in $\Sigma_8$ may be explained by the fact that COSMOS2015 tends to calibrate the KiDS galaxies to higher redshifts than the spectroscopic calibration data alone (see Fig.~\ref{fig:nz_compare_groups}), translating to lower $\Sigma_8$. The same reasoning does not explain why $\Sigma_8$ reduces further when the PAUS data is included, therefore this behaviour is most likely owed to the quality control (Eq.~\ref{eq:qc}). When the photo-$z$ data is combined with the spectroscopic compilation, it may result in a significantly different distribution of $| \langle z_{\rm cal} \rangle - \langle z_{\rm B} \rangle |$ for the SOM groupings, which has non-trivial implications for the gold sample selection and the derived cosmological constraints.

\section{Conclusion}
\label{sec:conclusion}

We apply the SOM redshift calibration technique \citep{Wright20a} to define and calibrate the redshift distributions of a new set of KiDS-1000 gold samples by adopting a new spectroscopic calibration sample. Compared to previous work by \citet{Hildebrandt21} we double the size of this calibration sample by adding more than ten additional spectroscopic campaigns such as C3R2 and DEVILS, which allows us to calibrate an additional \SI{9}{\percent} of the KiDS galaxies.

We take this one step further by enhancing the calibration sample with precision photometric redshifts from the PAU survey and COSMOS2015, maintaining a hierarchy that prefers spectroscopic over PAUS and COSMOS2015 redshifts to resolve duplicates in the three catalogues. The resulting KiDS gold sample increases by additional \SI{6}{\percent} and covers nearly \SI{98}{\percent} of all KiDS galaxies in the fifth tomographic bin.

When comparing these gold samples we find changes in the mean redshifts of up to $|\Delta z| = 0.026$ which originate from selection effects in the calibration sample. First, there are residual modifications to the redshift calibration of those KiDS sources that are found in both gold samples. These modifications are a direct consequence of changing the redshift distribution of the calibration sample when adding the photo-$z$. Second and most important, the selection of KiDS sources itself changes since the faint COSMOS2015 data allows us to calibrate additional galaxies at both low and high redshifts. These results highlight the importance of quantifying and calibrating potential method dependent redshift biases arising from selection effects, as has been shown in previous work. This requires sophisticated galaxy mock data with sufficient redshift coverage, realistic galaxy colours and accurate modelling of photometric and spectroscopic galaxy samples, in particular if one aims to study the impact of photo-$z$ outliers on the calibration sample.

In the second part of this study we perform a cosmic shear analysis using COSEBIs and find $S_8 = 0.748_{-0.025}^{+0.021}$%
 for our fiducial and $S_8 = 0.757_{-0.026}^{+0.016}$%
 for the photo-$z$-enhanced gold sample which is slightly lower than, but still in excellent agreement with, previous work on KiDS-1000 by \citet{Asgari21}. As part of our additional systematic testing we create a third gold sample that we calibrate using only the most secure spectra of our spectroscopic compilation (nQ $\geq 4$). We measure $S_8 = 0.766_{-0.023}^{+0.023}$%
 which corresponds to an increment of 0.5σ compared to the fiducial case. Our analysis speculates that this sample underestimates the true redshifts and overestimates $S_8$ due to an implicit selection introduced when limiting the calibration to the most secure spectroscopic redshifts \citep{Hartley20}. Therefore completeness and quality of the calibration sample can be optimised, this however requires realistic simulations to quantify the impact of this trade-off on the recovered redshift distributions.

Finally we analyse four KiDS gold samples which are all calibrated from different subsets of our extended calibration sample or using only photo-$z$ from PAUS and COSMOS2015. No matter how we calibrate the KiDS source galaxies, we find that all seven gold samples studied in this work scatter in the range of $S_8 = 0.743 \dots 0.766$ around our fiducial analysis. This further confirms previously reported tensions of KiDS with measurements of the CMB by \citet{Planck20} at 3.0σ to 3.6σ for a flat ΛCDM model.

In summary, there seems to be little benefit in using precision photometric redshifts for the SOM redshift calibration given the excellent coverage of the KiDS data by our spectroscopic compilation alone. The {\sl \mbox{spec-$z$}+\allowbreak PAUS+\allowbreak COS15} gold sample achieves, compared to our fiducial sample, a \SI{6}{\percent} improvement in terms of the number density, yet improvements on the cosmological constraints are marginal. Nevertheless, if the spectroscopic coverage is significantly lower or one wishes to target higher redshifts, photo-$z$ samples are a valuable source of complementary calibration data. However, the greater the dependence on photometric redshifts is, the more attention must be paid to redshift outliers to guarantee a good balance between statistical uncertainties and systematic biases in the redshift calibration. Given their challenging calibration requirements this is in particular true for the next generation, stage IV surveys such as Euclid \citep{Laureijs11} or the Vera C. Rubin Observatory Legacy Survey of Space and Time \citep[LSST,][]{Ivezic19}. Despite photo-$z$ outliers, photometric redshift samples have one significant advantage over spectroscopically selected samples: they achieve a much higher completeness, which can mitigate selection effects in the calibration sample and may improve the ability of the SOM to map out the full extent of colour-redshift-degeneracies.


\begin{acknowledgements}
  We acknowledge support from the European Research Council under grant numbers 770935 (JvdB, AHW, HH) and 647112 (MA, CH, TT).
  HH is also supported by a Heisenberg grant (Hi1495/5-1) of the Deutsche Forschungsgemeinschaft.
  MB is supported by the Polish National Science Center through grants no. 2020/38/E/ST9/00395, 2018/30/E/ST9/00698 and 2018/31/G/ST9/03388, and by the Polish Ministry of Science and Higher Education through grant DIR/WK/2018/12.
  CH is additionally supported by the Max Planck Society and the Alexander von Humboldt Foundation in the framework of the Max Planck-Humboldt Research Award endowed by the Federal Ministry of Education and Research.
  HYS acknowledges the support from CMS-CSST-2021-A01, NSFC of China under grant 11973070, the Shanghai Committee of Science and Technology grant No.19ZR1466600 and Key Research Program of Frontier Sciences, CAS, Grant No. ZDBS-LY-7013.
  We are grateful to Mara Salvato and the zCOSMOS team to give us early access to additional deep spectroscopic redshifts that were not available in the public domain,
  as well as to Dan Masters for sharing pre-release C3R2 DR3 data.
  This work is based on observations made with ESO Telescopes at the La Silla Paranal Observatory under programme IDs 100.A-0613, 102.A-0047, 179.A-2004, 177.A-3016, 177.A-3017, 177.A-3018, 298.A-5015, and on data products produced by the KiDS consortium.
  This work has made use of CosmoHub \citep{Carretero17}. CosmoHub has been developed by the Port d'Informació Científica (PIC), maintained through a collaboration of the Institut de Física d'Altes Energies (IFAE) and the Centro de Investigaciones Energéticas, Medioambientales y Tecnológicas (CIEMAT), and was partially funded by the "Plan Estatal de Investigación Científica y Técnica y de Innovación" program of the Spanish government.
  The figures in this work were produced with {\sc Matplotlib} \citep{Matplotlib} and {\sc GetDist} \citep{GetDist}.
  \\
  {\it Author Contributions.} All authors contributed to the development and writing of this paper. The authorship list is given in three groups: the lead authors (JLvdB, AHW, HH, MB, and MA), followed by two alphabetical groups. The first alphabetical group includes those who are key contributors to both the scientific analysis and the data products. The second group covers those who have either made a significant contribution to the data products or to the scientific analysis.
\end{acknowledgements}


\bibliographystyle{aa}
\bibliography{./references} 


\begin{appendix}

\section{Updated spectroscopic compilation}
\label{app:compilation}

\begin{table*}
    \centering
    \caption{
        Listing of the spectroscopic samples that enter the spectroscopic compilation summarising the sample sizes and mean redshifts. The values are calculated after removing duplicates between overlapping catalogues and matching with the KiDS imaging.}
    \label{tab:specz_samples}
    \begin{tabular}{lccl}
\hline\hline
Sample &           Count & $\langle z \rangle$ &                                             Reference \\
\hline
hCOSMOS      &    $\hphantom{00}503$ &             $0.308$ &                                    \citet{Damjanov18} \\
GAMA-G15Deep &  $\hphantom{0}1\,840$ &             $0.357$ &                              \citet{Kafle18,Driver22} \\
G10-COSMOS   &  $\hphantom{}14\,849$ &             $0.586$ &                                      \citet{Davies15} \\
ACES         &  $\hphantom{0}4\,233$ &             $0.593$ &                                      \citet{Cooper12} \\
OzDES        &    $\hphantom{00}930$ &             $0.638$ &                                      \citet{Lidman20} \\
DEVILS       &  $\hphantom{0}5\,222$ &             $0.682$ &                                      \citet{Davies18} \\
VIPERS       &  $\hphantom{0}2\,436$ &             $0.718$ &                                   \citet{Scodeggio18} \\
VVDS         &  $\hphantom{0}5\,190$ &             $0.737$ &                           \citet{LeFevre05,LeFevre13} \\
LEGA-C       &    $\hphantom{00}216$ &             $0.818$ &                                   \citet{vanderWel16} \\
DEEP2        &  $\hphantom{0}8\,564$ &             $0.962$ &                                      \citet{Newman13} \\
C3R2         &  $\hphantom{0}2\,512$ &             $0.980$ &       \citet{Masters17,Masters19,Euclid20,Stanford21} \\
DEIMOS       &  $\hphantom{0}1\,729$ &             $1.045$ &                                    \citet{Hasinger18} \\
GOODS        &  $\hphantom{0}1\,999$ &             $1.292$ &  \textit{ESO compilation of GOODS/CDF-S spectroscopy} \\
FMOS-COSMOS  &    $\hphantom{00}272$ &             $1.566$ &                                   \citet{Silverman15} \\
zCOSMOS      &  $\hphantom{0}1\,875$ &             $1.602$ &                \textit{private comm. from M. Salvato} \\
VUDS         &    $\hphantom{00}205$ &             $1.974$ &                                     \citet{LeFevre15} \\
VANDELS      &    $\hphantom{00}336$ &             $2.504$ &                                     \citet{Garilli21} \\
\hline
\end{tabular}

\end{table*}

Here we present the details of our extended and revised spectroscopic compilation discussed in Sect.~\ref{sec:data:spec}. It includes redshifts from surveys covering a number of deep extragalactic fields for which we have 9-band VST+VISTA photometry of similar quality as the main KiDS+VIKING data (`KiDZ'). In this paper we use 6 such fields shown in Fig.~\ref{fig:KiDZ}; three of them (COSMOS, CDF-S and VVDS-2h) are covered by a number of partly overlapping surveys and we merged them into our redshift calibration sample. In doing this, we translate the input redshift quality flags or assessments into our flag $\mathtt{NQ}$, with $\mathtt{NQ} \geq 4$ indicating the most secure spectroscopic redshifts (confidence \SI{99}{\percent} or more), while $3 \leq \mathtt{NQ} < 4$ indicate secure, but lower-confidence redshifts. The particular surveys are listed in Table~\ref{tab:specz_samples}, in the ascending order of their mean redshift determined after removing multiples and cross-matching with our imaging. The details of the selection and quality assignment for the particular samples are as follows:

\begin{itemize}

\item hCOSMOS \citep{Damjanov18}: we use all the galaxies from the published dataset and assign them redshift quality $\mathtt{NQ}=4$;

\item GAMA-G15Deep \citep{Kafle18,Driver22}: we select galaxies with input redshift quality $\mathtt{Z\_QUAL}\geq 3$ and with redshifts $z>0.001$ to avoid stellar contamination. We assign $\mathtt{NQ} = \mathtt{Z\_QUAL}$;

\item G10-COSMOS \citep{Davies15}: we use $\mathtt{Z\_BEST}$ as the redshift value and follow the recommended selection for galaxy spectroscopic redshifts, that is $\mathtt{Z\_BEST}>0.0001$, $\mathtt{Z\_USE}<3$ and $\mathtt{STAR\_GALAXY\_CLASS}=0$. As the redshifts provided in that compilation do not always have a quality flag that could be easily translated into our $\mathtt{NQ}$, we assign all these galaxies a specific flag $\mathtt{NQ}=3.5$;

\item ACES \citep{Cooper12}: we select redshifts with $\mathtt{Z\_QUALITY} \geq 3$ and with redshift errors below \SI{1}{\percent}. We assign $\mathtt{NQ} = \mathtt{Z\_QUALITY}$;

\item OzDES \citep{Lidman20}: we use two patches partly overlapping with KiDZ imaging (around CDF-S and around VVDS-2h). We select redshifts $z>0.002$ (stellar contamination removal) and require quality $\mathtt{qop} \in \lbrace3,4\rbrace$. We assign $\mathtt{NQ}=\mathtt{qop}$;

\item DEVILS \citep{Davies18}: we select spectroscopic redshifts only ($\mathtt{zBestType}=\mathtt{spec}$) and require the flags $\mathtt{starFlag}=0$, $\mathtt{mask}=0$, and $\mathtt{artefactFlag}=0$. As the DEVILS sample is a compilation including both own redshifts and external ones, we assign $\mathtt{NQ}=4$ if $\mathtt{zBestSource}=\mathtt{DEVILS}$ (redshift obtained from DEVILS observations) and $\mathtt{NQ}=3$ (external redshifts);
 
\item VIPERS \citep{Scodeggio18}: we require sources with the flag $2\leq\mathtt{zflg}<10$ or $22\leq\mathtt{zflg}<30$. We assign $\mathtt{NQ}=4$ if $3\leq\mathtt{zflg}<5$ or $23\leq\mathtt{zflg}<25$, and $\mathtt{NQ}=3$ otherwise;

\item VVDS \citep{LeFevre05,LeFevre13}: we join the WIDE, DEEP and UDEEP sub-samples, and use only sources with $\mathtt{ZFLAGS} \in \lbrace3,4,23,24\rbrace$. We assign $\mathtt{NQ}=4$ to these redshifts;

\item LEGA-C \citep{vanderWel16}: we select sources with the flag $\mathtt{f\_use}=1$ and assign $\mathtt{NQ}=4$ to them;

\item DEEP2 \citep{Newman13}: as in the previous KiDS papers \citep{Hildebrandt17,Hildebrandt20,Hildebrandt21}, we use sources from two equatorial fields (0226 \& 2330), require sources with $\mathtt{ZQUALITY} \geq 3$, and with redshift errors smaller than \SI{1}{\percent}. We assign $\mathtt{NQ}=\mathtt{Z\_QUALITY}$;

\begin{figure}
    \centering
    \includegraphics[width=\columnwidth,trim=0mm 5mm 0mm 5mm]{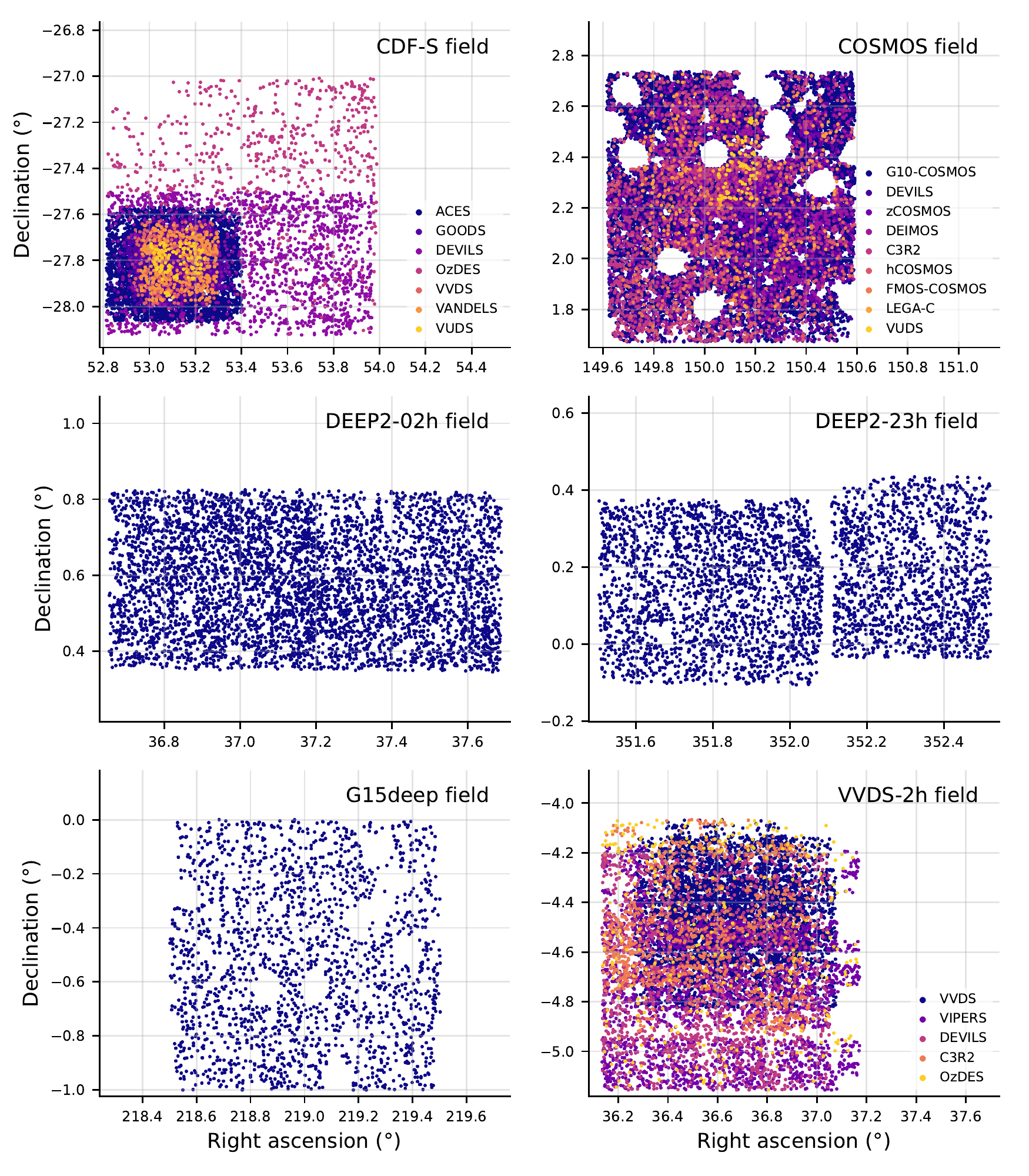}
    \caption{
        On-sky distribution of spectroscopic sources matched against the six KiDZ fields.}
    \label{fig:KiDZ}
\end{figure}

\begin{figure}
    \centering
    \includegraphics[width=\columnwidth,trim=0mm 5mm 0mm 5mm]{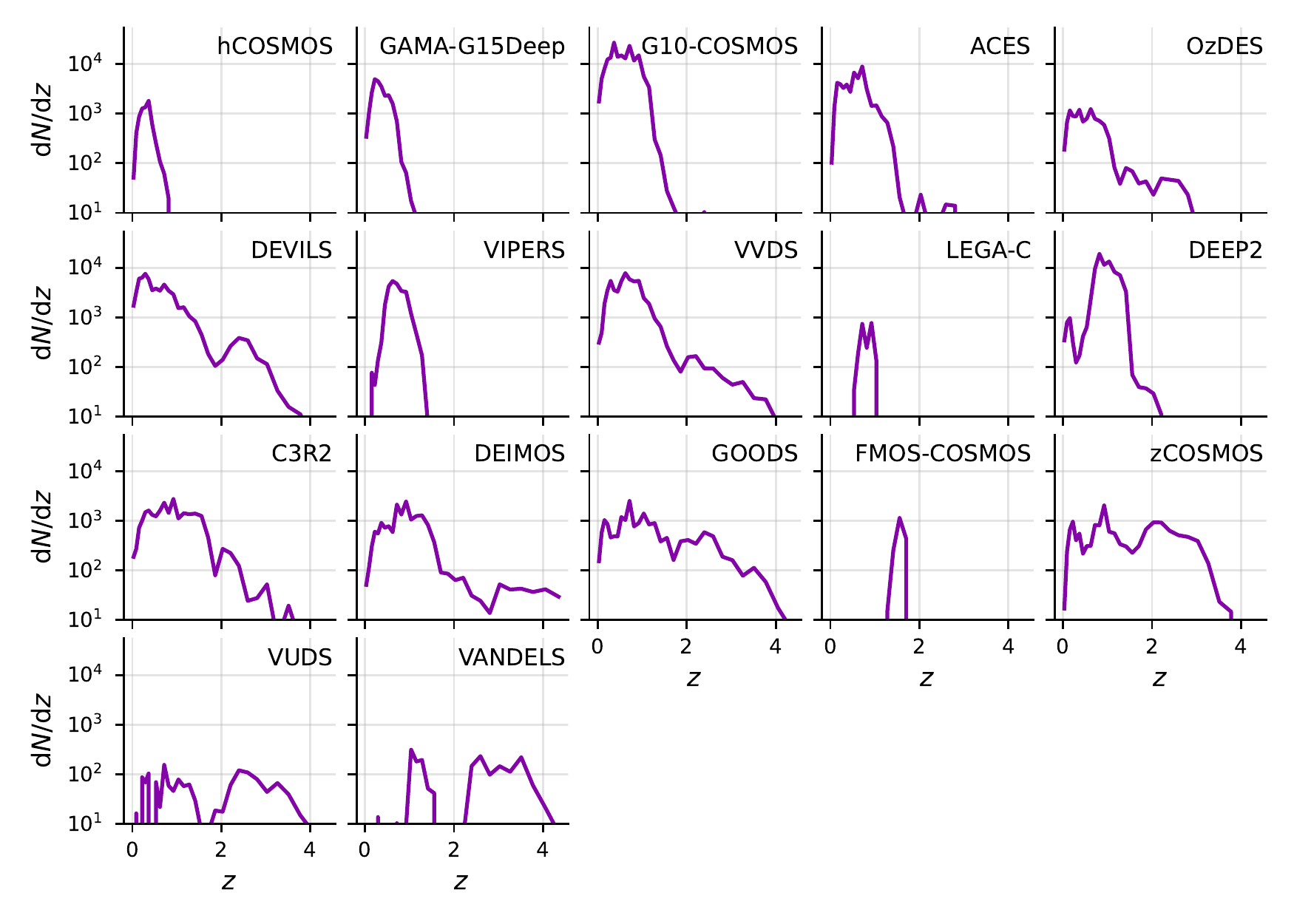}
    \caption{
        Redshift distributions of the spectroscopic samples that enter the spectroscopic compilation, ordered by mean redshift (Table~\ref{tab:specz_samples}). The distributions are calculated after removing duplicates between overlapping catalogues and matching with the KiDS imaging.}
    \label{fig:sample_nz}
\end{figure}

\item C3R2: we combine four public datasets, DR1 \citep{Masters17}, DR2 \citep{Masters19}, DR3 \citep{Stanford21}, and KMOS \citep{Euclid20} and exclude the Northern EGS field. We take sources with $\mathtt{QFLAG} \geq 3$ and assign $\mathtt{NQ}=\mathtt{QFLAG}$;

\item DEIMOS \citep{Hasinger18}: we require the quality flag $\mathtt{Q}=2$, while for assigning $\mathtt{NQ}$ we use the other flag provided, namely $\mathtt{NQ}=4$ for $\mathtt{Q_f} \in \lbrace 4,14 \rbrace$, $\mathtt{NQ}=3$ otherwise;

\item GOODS/CDF-S: we use a public ESO compilation of spectroscopy in the CDF-S field\footnote{Available from \url{https://www.eso.org/sci/activities/garching/projects/goods/MasterSpectroscopy.html}} \citep{Popesso09,Balestra10}. From the compilation, we choose secure redshifts (assigning $\mathtt{NQ}=4$ to them) and likely redshifts ($\mathtt{NQ}=3$) following the recommendations in the dataset description\footnote{\url{https://www.eso.org/sci/activities/garching/projects/goods/MASTERCAT_v3.0.dat}};

\item FMOS-COSMOS \citep{Silverman15}: we select sources with the quality flag $\mathtt{q\_z} \geq 2$ and assign $\mathtt{NQ}=4$ if $\mathtt{q\_z} =4$ or $\mathtt{NQ}=3$ otherwise;

\item zCOSMOS: we use this name for a proprietary compilation of various spectroscopic surveys in the COSMOS field, kindly provided to us by Mara Salvato, updated as of 01 September 2017. That dataset includes some of the surveys already included in our compilation, but it also provides redshifts from various other campaigns. We use the provided quality flag and select sources meeting the criteria $3 \leq \mathtt{Q\_f} \leq 5$ or $13 \leq \mathtt{Q\_f} \leq 15$ or $23 \leq \mathtt{Q\_f} \leq 25$ or $\mathtt{Q\_f} \in \lbrace 6, 10 \rbrace$, do not use sources from lower-confidence determinations (e.g. grism), and limit the redshifts to $\mathtt{z\_spec} > 0.002$ to avoid stellar contamination. We assign redshift quality as $\mathtt{NQ} = \min((\mathtt{Q_f}\!\mod 10),4)$;

\item VUDS \citep{LeFevre15}: we use sources with redshift flag $\mathtt{zflags}$ ending with $\lbrace 3, 4, 9\rbrace$ (reliability $\geq \SI{80}{\percent}$) and assign $\mathtt{NQ}=4$ if $3\leq \mathtt{zflags} <5$ or $13 \leq \mathtt{zflags} < 25$, otherwise $\mathtt{NQ}=3$;

\item VANDELS \citep{Garilli21}: we select sources for which $(\mathtt{zflg}\!\mod 10) \in \lbrace 2, 3, 4\rbrace$ and assign $\mathtt{NQ}=4$ if $(\mathtt{zflg}\!\mod 10) \in \lbrace 3, 4 \rbrace$ or $\mathtt{NQ}=3$ otherwise. The reassignment of the quality flags is motivated by the reportedly high redshift confidence of objects with flag values of 2 and 3.

\end{itemize}

When joining the above samples into one dataset, we remove duplicates both from overlapping surveys as well as within the individual ones. In the former case, if for a given source there are redshifts from different surveys available, we assign the most reliable measurement based on a specific `hierarchy'. Namely, we join the catalogues by cross-matching objects within \SI{1}{\arcsec} radius and apply the following order of preference:
\begin{itemize}
\item COSMOS field: G10-COSMOS > DEIMOS > hCOSMOS > VVDS > Lega-C > FMOS > VUDS > C3R2 > DEVILS > zCOSMOS;
\item CDF-S field: ACES > VANDELS > VVDS > VUDS > GOODS/CDF-S > DEVILS > OzDES;
\item VVDS-2h field: VIPERS > VVDS > C3R2 > DEVILS > OzDES.
\end{itemize}

For objects with multiple spectroscopic measurements within a particular survey, we either take the redshift with the highest quality flag or, if various entries for the same source have the same quality flag and the reported redshifts differ by no more than 0.005, we calculate the average. If the reported redshifts have the same quality flag but differ by more than 0.005, we exclude such a source from the compilation.

\section{SOM quality control}
\label{app:SOMqc}

The original quality control (Eq.~\ref{eq:qc}) was calibrated for the spectroscopic calibration data of \citet{Hildebrandt21} with $\sigma_{\rm mad} = \mathrm{nMAD}\left( \langle z_{\rm cal} \rangle - \langle z_{\rm B} \rangle \right) \approx 0.12$. Since this criterion may not be optimal for our new calibration dataset, we would in principle need to recalibrate the SOM cell rejection threshold to obtain the optimal trade-off between redshift bias and the size of the gold sample. It is, however, very challenging to simulate the photo-$z$ errors and success rates of our new calibration dataset and therefore we choose to keep the threshold fixed for all gold sets.

\begin{figure}
    \centering
    \includegraphics[width=\columnwidth]{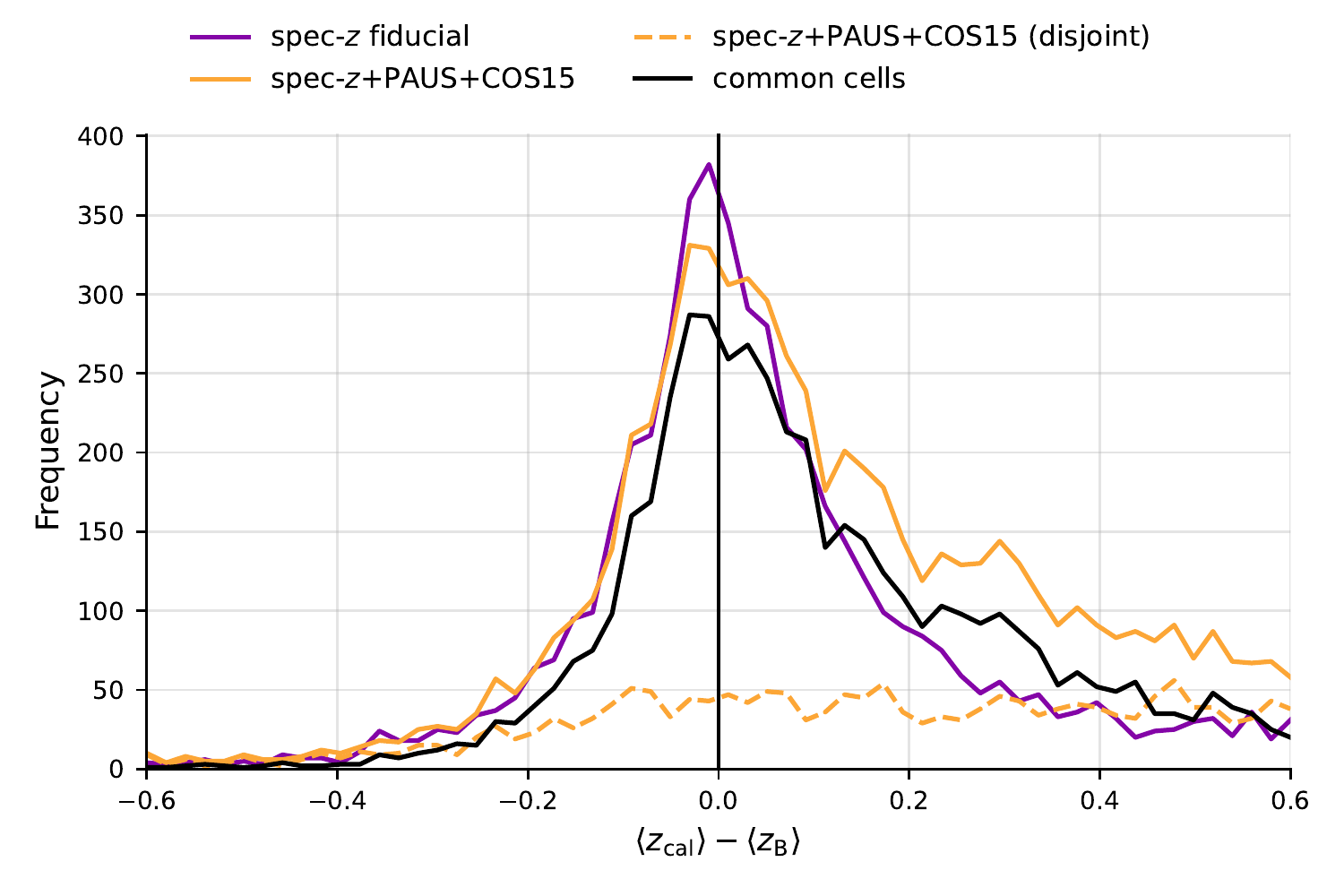}
    \caption{
        Comparison of the mean calibration sample redshift and KiDS 9-band photo-$z$ in each SOM cell within the fiducial $\pm 5\sigma_{\rm mad}$ interval. The coloured solid lines represent the fiducial (purple) and the full (orange) gold sample. The black line is the subset of SOM cells that belong to both gold samples, whereas the dashed orange line represents those cells that are occupied only by data from COSMOS2015.}
    \label{fig:SOMcellbias}
\end{figure}

Nevertheless, the addition of COSMOS2015 to the calibration data has a significant impact on the distribution of $\langle z_{\rm cal} \rangle - \langle z_{\rm B} \rangle$ for the fully trained cells of the SOM (Fig.~\ref{fig:SOMcellbias}). While the distribution is almost symmetrical for the case of {\sl \mbox{spec-$z$} fiducial}, it is skewed to positive values for {\sl \mbox{spec-$z$}+\allowbreak PAUS+\allowbreak COS15}. The effect is more distinct for those cells that are occupied exclusively by the COSMOS2015 data than for those that exist in both gold samples (compare the orange dotted and black lines in Fig.~\ref{fig:SOMcellbias}). This is a clear indication that much of the additional calibration data are faint objects for which the KiDS 9-band photo-$z$ catastrophically underestimates the redshift obtained from the much deeper COSMOS2015 data. This effect is responsible for the reduced representation fraction in the first tomographic bin, as discussed in Sect.~\ref{sec:newgold_mix}.

\section{Differences from the original KiDS-1000 analysis}
\label{app:K1000diff}

\begin{table*}
    \centering
    \caption{
        Summary of the main cosmological parameter constraints (best fit and 68th-percentile PJ-HPD) from COSEBIs for \citet{Asgari21} and our reanalysis with and without the corrected $m$-bias values.}
    \renewcommand{\arraystretch}{1.33}
    \label{tab:params_A21}
    \begin{tabular}{lcccccc}
\hline\hline
Sample & $\chi^2$ &                           $A_{\rm IA}$ &           $\Omega_{\rm m}$ &                 $\sigma_8$ &                      $S_8$ &                 $\Sigma_8$ \\
\hline
Asgari et al. (2021) &     82.2 &  $\hphantom{-}0.264_{-0.336}^{+0.424}$ &  $0.246_{-0.060}^{+0.101}$ &  $0.838_{-0.141}^{+0.140}$ &  $0.759_{-0.021}^{+0.024}$ &  $0.751_{-0.016}^{+0.024}$ \\
Asgari reanalysis    &     83.8 &  $\hphantom{-}0.243_{-0.424}^{+0.370}$ &  $0.240_{-0.059}^{+0.101}$ &  $0.846_{-0.146}^{+0.146}$ &  $0.758_{-0.018}^{+0.029}$ &  $0.749_{-0.013}^{+0.025}$ \\
Asgari new $m$-bias  &     83.2 &  $\hphantom{-}0.233_{-0.388}^{+0.406}$ &  $0.242_{-0.060}^{+0.104}$ &  $0.844_{-0.165}^{+0.114}$ &  $0.757_{-0.023}^{+0.022}$ &  $0.749_{-0.017}^{+0.024}$ \\
\hline
\end{tabular}

    \renewcommand{\arraystretch}{1.0}
    \tablefoot{
        Additionally shown are the $\chi^2$ values (for $70.5$ effective degrees of freedom) and $\Sigma_8 = \sigma_8 (\Omega_{\rm m} / 0.3)^\alpha$ calculated for \protect$\alpha_{\rm fid} = 0.55$%
, determined from a fit to the posterior samples of the fiducial chain.}
\end{table*}

The major difference between the analysis of the original KiDS-1000 data and this work is that our redshift calibration strategy focuses exclusively on improving the SOM redshifts. We omit revisiting the KiDS clustering redshifts, which \citet{Hildebrandt21} used as a validation for the fiducial SOM redshift distributions. We note, however, that the additional calibration data is only beneficial for the SOM calibration, since such inhomogeneous data, in particular photometric redshifts, cannot be applied easily in clustering redshifts.

Furthermore there are small differences in the cosmological analysis compared to \citetalias{Asgari21}, which are mostly related to small differences in the COSEBI data vector. We find that these differences originate from using a newer version of {\sc TreeCorr}, however it has no impact on our cosmological constraints which agree within the expected statistical variance of the sampler. Differences in the measured COSEBIs also propagate to the covariance matrix, resulting in sub-percent differences between our covariance and the one used by \citetalias{Asgari21}. Finally, we assume \SI{99}{\percent} correlation between $m$-bias values per tomographic bin instead of \SI{100}{\percent} (perfect correlation) as in \citetalias{Asgari21}. The effect of this choice is negligible compared to those introduced by the variations in the COSEBI data vector.

We also test whether the correction of the $m$-bias (Table~\ref{tab:mbias}) has a significant impact on the cosmological constraints in \citetalias{Asgari21} by analysing the same data (`Asgari new $m$-bias') with otherwise identical parameters (except for the differences summarised above). Our results show that the recovered primary cosmological parameters, in particular $S_8 = 0.757_{-0.023}^{+0.022}$%
, are almost identical with those of our reanalysis of the KiDS-1000 data, performed with the old $m$-bias values (Table~\ref{tab:params_A21}). Given that the uncertainties of the $m$-biases are unchanged and the central value changes by less than 0.5σ in any of the tomographic bins, these results are as expected and have no further implication for the conclusions of \citetalias{Asgari21}.

\section{Marginal parameter constraints}

Table~\ref{tab:params_marginal} lists an alternative representation of the parameter constraints of Table~\ref{tab:params}, showing the maximum of the marginal distribution and the associated 68th-percentile instead of the best fit and the 68th-percentile PJ-HPD.

\begin{table*}
    \centering
    \caption{
        Summary of the main cosmological parameter constraints (maximum of the marginal distribution and the associated 68th-percentile) from COSEBIs for all gold samples and the comparison to \citet{Asgari21} and Planck legacy (TT, TE, EE + lowE).}
    \renewcommand{\arraystretch}{1.33}
    \label{tab:params_marginal}
    \begin{tabular}{lccccc}
\hline\hline
Sample &                           $A_{\rm IA}$ &           $\Omega_{\rm m}$ &                 $\sigma_8$ &                      $S_8$ &                 $\Sigma_8$ \\
\hline
spec-$z$ fiducial             &  $\hphantom{-}0.394_{-0.387}^{+0.328}$ &  $0.278_{-0.084}^{+0.097}$ &  $0.734_{-0.129}^{+0.134}$ &  $0.745_{-0.026}^{+0.020}$ &  $0.740_{-0.019}^{+0.019}$ \\
spec-$z$+PAUS                 &  $\hphantom{-}0.108_{-0.441}^{+0.347}$ &  $0.238_{-0.066}^{+0.101}$ &  $0.800_{-0.159}^{+0.118}$ &  $0.745_{-0.025}^{+0.019}$ &  $0.736_{-0.019}^{+0.020}$ \\
spec-$z$+PAUS+COS15           &  $\hphantom{-}0.440_{-0.465}^{+0.328}$ &  $0.206_{-0.057}^{+0.109}$ &  $0.843_{-0.163}^{+0.152}$ &  $0.750_{-0.024}^{+0.019}$ &  $0.739_{-0.021}^{+0.018}$ \\
spec-$z$ $\mathrm{nQ} \geq 4$ &  $\hphantom{-}0.088_{-0.378}^{+0.395}$ &  $0.203_{-0.050}^{+0.080}$ &  $0.877_{-0.138}^{+0.134}$ &  $0.758_{-0.022}^{+0.019}$ &  $0.746_{-0.019}^{+0.019}$ \\
only PAUS                     &  $\hphantom{-}0.078_{-0.488}^{+0.367}$ &  $0.210_{-0.057}^{+0.112}$ &  $0.879_{-0.202}^{+0.112}$ &  $0.753_{-0.025}^{+0.021}$ &  $0.740_{-0.017}^{+0.023}$ \\
only COS15                    &  $\hphantom{-}0.345_{-0.452}^{+0.324}$ &  $0.245_{-0.072}^{+0.137}$ &  $0.672_{-0.100}^{+0.204}$ &  $0.736_{-0.023}^{+0.024}$ &  $0.733_{-0.019}^{+0.020}$ \\
only PAUS+COS15               &  $\hphantom{-}0.392_{-0.476}^{+0.349}$ &  $0.264_{-0.099}^{+0.111}$ &  $0.743_{-0.166}^{+0.149}$ &  $0.737_{-0.026}^{+0.021}$ &  $0.730_{-0.020}^{+0.020}$ \\
Asgari et al. (2021)          &  $\hphantom{-}0.384_{-0.408}^{+0.360}$ &  $0.254_{-0.076}^{+0.088}$ &  $0.773_{-0.125}^{+0.145}$ &  $0.758_{-0.026}^{+0.017}$ &  $0.753_{-0.023}^{+0.015}$ \\
Planck legacy                 &                                    --- &  $0.316_{-0.008}^{+0.009}$ &  $0.813_{-0.008}^{+0.007}$ &  $0.836_{-0.017}^{+0.016}$ &  $0.838_{-0.018}^{+0.017}$ \\
\hline
\end{tabular}

    \renewcommand{\arraystretch}{1.0}
    \tablefoot{
        Additionally shown is $\Sigma_8 = \sigma_8 (\Omega_{\rm m} / 0.3)^\alpha$ calculated for \protect$\alpha_{\rm fid} = 0.55$%
, determined from a fit to the posterior samples of the fiducial chain.}
\end{table*}

\section{Goodness-of-fit}

\begin{figure*}
    \centering
    \includegraphics[width=0.9\textwidth]{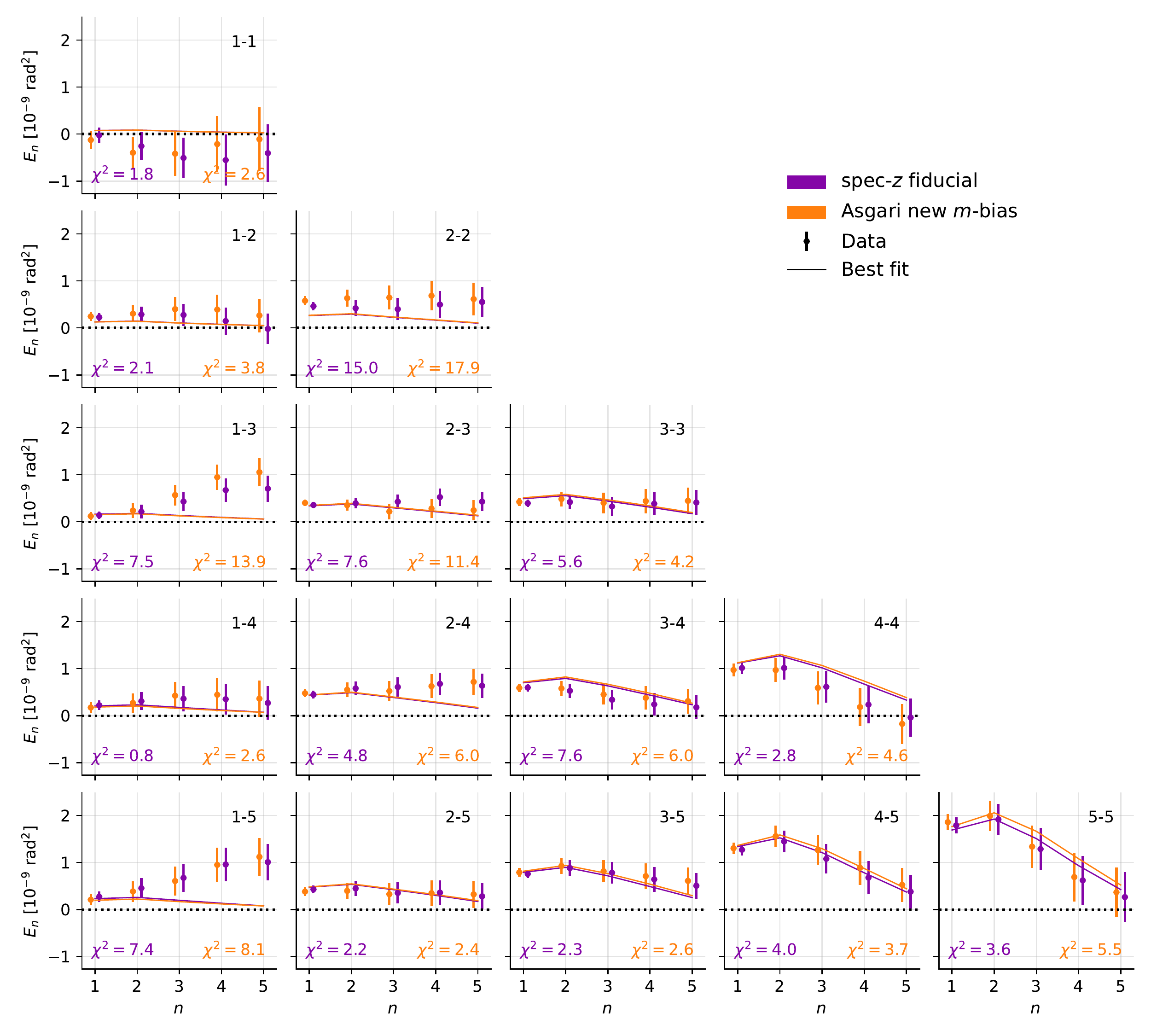}
    \caption{
        Comparison of the COSEBI data vector and the best fit model for {\sl \mbox{spec-$z$} fiducial} and the reanalysis of \citet{Asgari21} with corrected $m$-bias. The triangular arrangement of all 15 combinations of tomographic bins are labelled with their individual $\chi^2$-values. The COSEBIs modes are significantly correlated, therefore their goodness-of-fit cannot be estimated by eye.}
    \label{fig:bestfit}
\end{figure*}

We find that the cosmological model fits to the new gold samples achieve a better goodness-of-fit ($\chi^2$ between 62 and 74, Table~\ref{tab:params}) than those to the original KiDS-1000 gold sample with $\chi^2 \approx 83$ as reference. Since our methodology is identical in both cases, these differences stem from changes of the data vector that propagate also into the covariance matrix. We compare in particular the COSEBI data vectors with their respective best fit models for {\sl \mbox{spec-$z$} fiducial} and Asgari new $m$-bias (App.~\ref{app:K1000diff}) in Fig.~\ref{fig:bestfit}. Although a comparison by eye of model and data is difficult due to the strong correlation between the different COSEBI modes, the most significant changes in the data vector occur between modes that involve bin 1 and 2. In contrast to this the fitted models are very close in the first two bins and begin to deviate with increasing redshift.

These changes propagate to the $\chi^2$, which becomes apparent when summing the individual $\chi^2$-values in each panel of Fig.~\ref{fig:bestfit} for the two left-most columns. While the sum over the remaining three columns (dominated by bin 3, 4, and 5) is almost identical for {\sl \mbox{spec-$z$} fiducial} and Asgari new $m$-bias, the first two columns fully account for the smaller goodness-of-fit of $\Delta \chi^2 \approx 20$. Since changing the $m$-bias seems to have a rather small impact on the $\chi^2$ (see Table~\ref{tab:params_A21}), this behaviour is most likely linked to the changes in the gold sample selection and redshift distributions, which are most significant for the first two bins. \citetalias{Asgari21} report tensions between parameter constraints obtained from using only the second bin (including its cross-correlations) and the remainder of the data vector. A reduction of the $\chi^2$-value alone does not allow any conclusions on whether the improved calibration sample has an impact on the internal consistency of the cosmological analysis.

\end{appendix}


\end{document}